\documentclass[a4paper,fleqn]{cas-dc}

\usepackage[numbers,sort&compress]{natbib}
\begin{document}
\shorttitle{Quantum entanglement dynamics}
\shortauthors{Z.-B. Chen}

\title [mode = title]{Quantum entanglement dynamics of spacetime and matter}

\author{Zeng-Bing Chen}[type=editor,
                        auid=000,bioid=1,
                        orcid=0000-0002-9911-9356]
\ead{zbchen@nju.edu.cn}

\address{National Laboratory of Solid State Microstructures and School of Physics,
Nanjing University, Nanjing 210093, China}

\begin{abstract}
It was known long ago that quantum theory and general relativity are 
in sharp conflict in their foundations. Their
fundamental inconsistencies render a consistent theory of quantum gravity the
most challenging problem in physics. Here we propose an information-complete
quantum field theory (ICQFT), which describes elementary fermions, their gauge
fields, and gravity (together, called the trinary fields) as an elementary
trinity without any conceptual inconsistency of existing theories. The ICQFT 
unifies matter and spacetime (gravity) as information via spacetime-matter
entanglement, which encodes complete physical predictions
of the theory and leads to a compelling solution to the problem of time. We consider two 
particular forms of spacetime-matter entangled states
and their physical consequences. One of them results in a universal relation
between entanglement entropy and geometry (area and volume), allowing us to determine 
the cosmological constant term in the classical Einstein equation. Based on a quantum-information 
definition of dark energy, our Universe is not strictly holographic. We
predict the interior quantum state of a Schwarzschild black hole to be
maximally information-complete. As a concrete quantum formulation of gravity
coupled with matter, the ICQFT is quantum entanglement dynamics for
spacetime and matter and eliminates the conceptual obstacles of existing
quantum gravity theory.

\end{abstract}


\maketitle

\section{Introduction}

Quantum theory---quantum mechanics and quantum field theory (QFT)---and general
relativity are two pillars of our current physics and deeply impact even our
daily life. The achievements motivated by either of the two pillars are
remarkable. However, it was recognized long ago that quantum theory and
general relativity are in sharp conflict in their foundations as summarized by
Thiemann in a beautiful review \cite{Thiemann}. Einstein's equation relates
the geometry of spacetime and the energy-momentum tensor of matter. In
Thiemann's terminology, on one hand, the classical-quantum inconsistency means
that, while the matter fields are well described by the Standard Model in flat
spacetime, the geometry of spacetime is described by the classical Einstein
equation. On the other hand, general relativity results in the unavoidable
existence of spacetime singularities, where all laws of physics are doomed to
fail. Such an instability of spacetime and matter implies the internal
inconsistency of general relativity. At the same time, conventional QFT
suffers from the notorious infrared and ultraviolet singularities
(divergences). Although the divergences can be \textquotedblleft get rid
of\textquotedblright\ by renormalization as in the Standard Model,
renormalization fails when it applies to general relativity.

The fundamental inconsistencies mentioned above motivate a long march of
quantizing gravity---\textquotedblleft quantum theory's last
challenge\textquotedblright\ \cite{AC-nature}. Among various existing
approaches to quantum gravity, loop quantum gravity
\cite{Thiemann,quanA1,quanA2,quanA3,loop25,Rovelli-book,Thiemann-reg,Chiou} is
very impressive for, among others, its prediction of discrete structure of
spacetime and the entropy counting of black holes. However, almost all, 
if not all, existing methods to quantum gravity tacitly
assume, explicitly or implicitly, the completeness of conventional quantum
theory. \textit{Logically, it is totally possible that the fundamental
inconsistencies of our current theories could be caused by the incompleteness
of quantum theory}. The debate on the real meaning of quantum states
\cite{EPR,Bohr,PBR} and on the quantum measurement problem
\cite{vonN-book,WZ-book,Zurek,Sun} occupies the whole history of quantum
theory. Notice that conventional QFT in curved spacetime has
its own interpretational problems, e.g., the black-hole information paradox
\cite{Susskind-BH} and the physical meaning on the usual concept of particles
\cite{Unruh,QFT-cs,Unruh-RMP}. These interpretational difficulties of quantum
theory motivate various interpretations \cite{Intp-book}, or understanding
quantum theory from different angles \cite{Fuchs,i-causality,Colbeck-Renner}.

Yet, these interpretations or fresh understanding seldom
challenges the completeness of quantum theory. The most serious challenge stems from the famous
Einstein-Podolsky-Rosen paper \cite{EPR} questioning the completeness of
current quantum description against local realism. The follow-up discover of
Bell's inequalities \cite{Bell} and their various experimental tests give us
an impression that quantum mechanics wins against the Einstein-Podolsky-Rosen
argument. The interpretation on violations of Bell's inequalities as quantum
nonlocality was questioned from the many-worlds picture \cite{Tipler}.

Recently, we took a totally different way of thinking. We suggested an
information-complete quantum theory (ICQT) by assuming that quantum states
represent an information-complete code of any possible information that one
might access to a physical system \cite{ICQT}. The key to this development is
the \textit{information-completeness in the trinary picture}. A single, free
physical system in our conventional sense is excluded from the outset by the
ICQT as it is simply meaningless for acquiring information, which must be
accessed via interaction (entanglement). The two-party (a physical system
$\mathcal{S}$ plus its measurement apparatus $\mathcal{A}$) picture as used in
current quantum mechanics was argued to be information-incomplete. To fulfill
the information-completeness such that \textit{any information must be carried
or acquired by certain quantum system}, one has to adapt a trinary
description, in which the third system, called the \textquotedblleft
programming\textquotedblright\ system (system $\mathcal{P}$), has to be
introduced. Then the whole system $\mathcal{P}$-$\mathcal{SA}$ (the
\textquotedblleft trinity\textquotedblright) possesses a particular
self-defining quantum structure without the usual measurement postulate. But
in the context of conventional quantum mechanics, one could introduce more
programming systems $\mathcal{P}^{\prime}$, $\mathcal{P}^{\prime\prime}$... 
to programe $\mathcal{P}$-$\mathcal{SA}%
$, $\mathcal{P}^{\prime}$-$\mathcal{(P}$-$\mathcal{SA)}$...,
known in the usual quantum measurement model as the von Neumann chain, 
which is mathematically unlimited. As we
argued previously, if $\mathcal{P}$ is spacetime being a physical quantum
system, the von Neumann chain is terminated as there is no spacetime beyond
spacetime. Now we immediately see that, to arrive at an information-complete
and self-defining quantum structure, spacetime must be a quantized physical
system, as well as the programming system. Meanwhile, general relativity tells
us that spacetime is indeed a physical system and the same thing as gravity.

Now a fascinating thing happens here. On one hand, a genuine ICQT requires
that spacetime/gravity must be quantized and plays a very specific role in its
own formulation. On the other hand, we do have the most remarkable trinity of
nature---matter fermions, their gauge fields, and gravity (spacetime); the
role of the Higgs field will be considered elsewhere \cite{gGUT}. In the
present work, we generalize the idea of the ICQT and present an
information-complete QFT (ICQFT), which describes elementary
fermions, their gauge fields, and gravity as an indivisible quantum trinity.
By its very construction, the ICQFT provides a coherent picture and conceptual
framework of unifying matter and spacetime (gravity) as information via
spacetime-matter entanglement. Such a quantum information (or, entanglement)
dynamics of spacetime and matter represents thus a candidate unifying quantum
theory and general gravity into a single, consistent theory. The formulation
gives up the probability description of current quantum mechanics and does not
need the vague concepts such as observers and wave-function collapse.
\textit{The theory describes a self-defining or self-explaining Universe that
is genuinely quantum; there is no room for any classical systems or concepts}.
We consider some applications of the ICQFT. First, we give a compelling
solution to the problem of time, well-known in
quantum gravity community, and a quantum information definition of dark
energy. Second, a particular form of spacetime-matter entanglement allows us
to give a correct classical limit (i.e., the classical Einstein equation).
Then we conjecture a spacetime-matter state of the Universe, permiting a
natural generalization of the holographic relation, in which an extra term is
argued to be related to dark energy and the cosmological constant term in
Einstein's equation. Finally, we predict the interior quantum state of a
Schwarzschild black hole to be maximally information-complete. Central to our
development is the idea of spacetime-matter entanglement that encodes complete
physical predictions of the theory. In our formalism, \textit{entanglement is
thus universal just like that gravity is universal; the universal entanglement
is the glue of spacetime and matter and thus the building block of the world}.

\section{Information-complete quantum fields and the state-dynamics
postulate}

\begin{figure}
\centering
\includegraphics[scale=0.45]{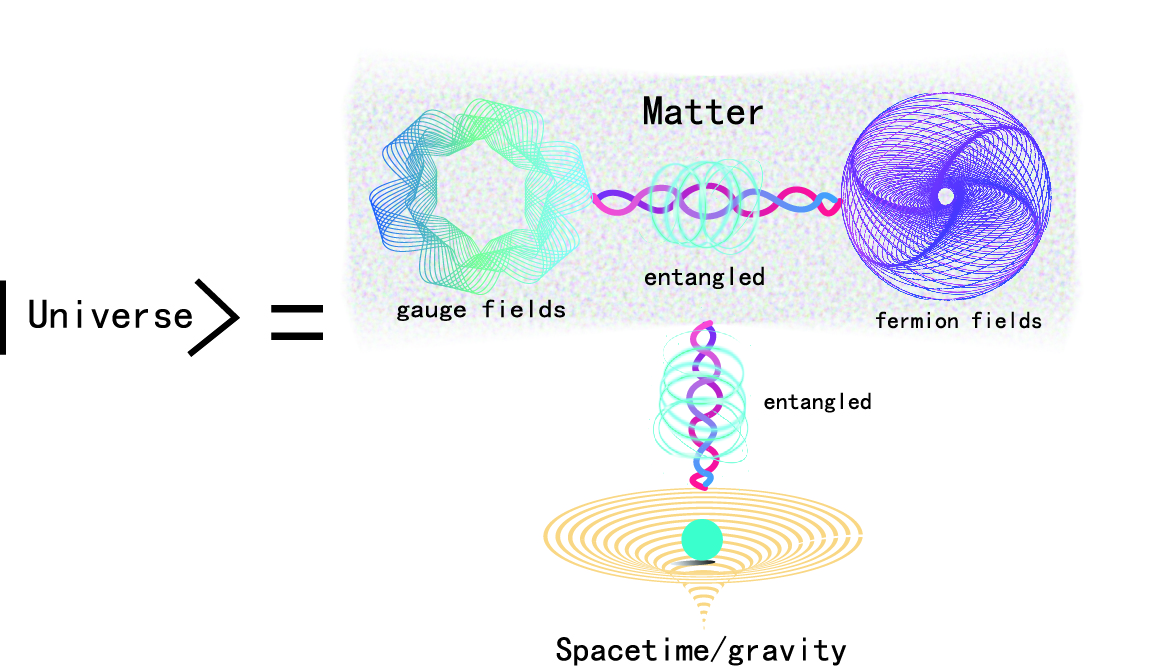}
\caption{\textbf{The fundamental trinity of nature}. Here spacetime-matter entanglement ``glues'' 
spacetime and matter (matter fermions and their gauge fields) as an indivisible trinity, 
encodes information-complete physical predictions of the world, and is as 
universal as universal gravitation.}
\label{Tri}
\end{figure}

In the present case of the ICQFT, an elementary fermion (e.g., a Dirac
electron) field and its corresponding gauge field are called as system
$\mathcal{S}$ and system $\mathcal{A}$, respectively; $\mathcal{SA}$ together
as matter fields. The gravitational field (i.e., quantized spacetime) is the
programming\ system (system $\mathcal{P}$). The trinary description (see Fig.~\ref{Tri}) then
corresponds to a dual entanglement pattern among the three systems: The matter
fields ($\mathcal{S}$ and $\mathcal{A}$ together) and gravity are mutually
defined by interacting and entangling each other; the fermion field
($\mathcal{S}$) and the gauge field ($\mathcal{A}$), both programmed by
gravity, are likewise entangled and mutually defined. Here spacetime plays a
role of quantum memory that stores or encodes all entanglement patterns for
the fermion field and its gauge field. Thus, \textit{in the ICQFT the
viewpoint on spacetime and matter is dramatically different from our previous
picture. Neither spacetime nor matter is an isolated entity; they must be
described as a trinity and entangled in the dual form to make sense for
acquiring information}. In this sense, the ICQFT is quantum entanglement
dynamics of the trinity, in which quantized spacetime plays a unique role for
formulating the theory.

To illustrate the basic idea, for concreteness we only consider the
gravitational field interacting with two kinds of matter fields: the fermion
field $\hat{\psi}(x)=\hat{\psi}(\mathbf{x},t)$ of a Dirac particle (with mass
$m$ and charge $q$) and the electromagnetic field $\hat{F}_{\mu\nu}%
=\partial_{\mu}\hat{A}_{\nu}-\partial_{\nu}\hat{A}_{\mu}$, where $\hat{A}%
_{\mu}$ is the electromagnetic potential vector. The generalization to
non-Abelian gauge fields is straightforward. Here we adapt notions as used in
Rovelli's book \cite{Rovelli-book}. A spacetime coordinate $x=(x^{\mu})$ with
$\mu,\nu,...=0,1,2,3$ being spacetime tangent indices. The gravity is
described by the tetrad field $e_{\mu}^{I}(x)$, which relates to the usual
metric tensor $g_{\mu\nu}$ by $g_{\mu\nu}(x)=\eta_{IJ}e_{\mu}^{I}(x)e_{\nu
}^{J}(x)$. Indices $I,J,...$ label the Minkowski vectors and the Minkowski
metric $\eta_{IJ}$ has signature $\left[  -,+,+,+\right]  $. The total action
of the trinary fields is $S(\hat{e},\hat{\omega};\hat{A},\hat{\psi
})=S_{\mathrm{G}}(\hat{e},\hat{\omega})+S_{\mathrm{M}\text{\textrm{-}%
}\mathrm{G}}(\hat{e},\hat{\omega};\hat{A},\hat{\psi})$, where $S_{\mathrm{M}%
}(\hat{e},\hat{\omega};\hat{A},\hat{\psi})=S_{\mathrm{Dirac}}(\hat{e}%
,\hat{\omega};\hat{\psi},\hat{A})+S_{\mathrm{EM}}(\hat{e};\hat{A})$. Here we
only write down explicitly the action for Dirac's field to introduce
notations:
\begin{equation}
S_{\mathrm{Dirac}}\mathcal{=}\frac{1}{2}\int dx^{4}\hat{e}\bar{\psi}\left[
\gamma^{I}\hat{e}_{I}^{\mu}iD_{\mu}-m\right]  \hat{\psi}+\mathrm{H.C}.
\label{DEM}%
\end{equation}
where $\bar{\psi}=\hat{\psi}^{\dag}\gamma^{0}$, $D_{\mu}=\partial_{\mu}%
+\hat{\omega}_{\mu J}^{I}L_{\ I}^{J}-iq\hat{A}_{\mu}$, $\hat{e}$ is the
determinant of $\hat{e}_{I}^{\mu}$, $\gamma$'s are the Dirac matrices,
$\hat{\omega}$ is the spin connection, and $L_{\ I}^{J}$ are the generators of
the Lorentz algebra.

Due to the presence of quantized spacetime, it is convenient to work in the Hamiltonian formalism
\cite{Rovelli-book,Thiemann-reg,Chiou}, which is better established in loop
quantum gravity. There, the dynamical variables, in terms of $\hat{e}_{I}%
^{\mu}$ and $\hat{\omega}_{\mu J}^{I}$, are the Ashtekar-Barbero
\cite{Ashtekar,Barbero} connection $\mathcal{\hat{B}}_{a}^{i}({\tau})$
[defined on a three-dimensional surface without boundaries; $a,b,...=1,2,3$
are spatial indices and $i,j,...=1,2,3$ take values in the Lie algebra
$su(2)$] and the \textquotedblleft gravitational electric
field\textquotedblright\ $\mathcal{\hat{E}}_{b}^{j}({\tau})$, which is
the $i8\pi\gamma G$ ($\gamma$: the Immirzi parameter \cite{Immirzi}) times
momentum conjugate to $\mathcal{\hat{B}}_{a}^{i}({\tau})$.

According to the ICQT and as we pointed out above, we must describe the
trinity ($\hat{\psi}$, $\hat{A}$, and $\mathcal{\hat{B}}$) as a single,
information-complete physical system. We can formally write the entangled
state of the trinity as $\left\vert \mathcal{B},\psi,A\right\rangle $, which
always allows a $\mathcal{P}$-$\mathcal{AS}$ Schmidt decomposition with two
orthogonal bases of the matter field ($\hat{\psi}$,$\hat{A}$) sector and the
gravitational field ($\mathcal{\hat{B}}$) sector. This spacetime-matter
entanglement in the Schmidt form encodes complete information of the trinity
such that matter and spacetime are mutually defined. As programmed by each
state of gravity in the $\mathcal{P}$-$\mathcal{AS}$ Schmidt decomposition, the $\hat{\psi}$-$\hat{A}$
entanglement encodes complete information of the matter field sector such that
the fermion field $\hat{\psi}$ and the gauge field $\hat{A}$ are mutually defined.

Thus, in the present field-theoretical case, the trinity is also entangled in
dual form \cite{ICQT}, to be given explicitly below. Here only pure-state
entanglement appears in our description and is uniquely quantified by the
usual entanglement entropy \cite{pureEE1,pureEE2}. The property of the
pure-state entanglement entropy indicates that, while states for each field is
relative, their information encoded in dual entanglement is invariant under
the changes of \textquotedblleft local\textquotedblright\ (single-field)
bases, i.e., under any unitary transformations upon states of a single field.
This is in a perfect analog to the spirit of general relativity, in which
physical observables must be invariant under general coordinate
transformations. Now let us state the key point of the ICQFT---\textit{quantum
relationalism}: Complete information (namely, all
physical predictions) of the trinary fields (fermions, their gauge fields, and
gravity) is encoded in dual entanglement; fields involved in the
dual-entanglement structure should be mutually defined, as we specified above,
to obey the information-completeness.

According to our previous definition of information-complete physical systems
\cite{ICQT}, information-complete field states and field operators can only be
defined in the Hilbert spaces of the matter field ($\hat{\psi}$,$\hat{A}$)
sector and the gravitational field ($\mathcal{\hat{B}}$) sector. The
$\hat{\psi}$-$\hat{A}$ entanglement, programmed by each state in the
orthogonal basis of the gravitational field sector, encodes information for
information-incomplete field in the Hilbert space of $\hat{\psi}$ or $\hat{A}%
$. Thus, the ICQFT describes nature with the basic trinity that is in a
specific entanglement structure. The physical significance of our current
understanding on matter fermion fields and gauge fields is completely changed
in the ICQFT: Either a fermion field or its gauge field alone loses its
physical significance and cannot be regarded as isolated, physical
(information-complete) entities; only jointly they define spacetime and can be
described as an information-complete physical entity. This immediately
explains, to be shown below, the occurrence of the black-hole information
paradox \cite{Susskind-BH} as an unavoidable consequence of the
conventional information-incomplete description.

What is the dynamics leading to the proposed entanglement structure of the
whole system? If we include all gauge and fermion fields into the total action
$S(\mathcal{\hat{B}};\hat{A}...,\hat{\psi}...)$, then the ICQFT, by
definition, is a theory about the whole Universe. All predictions of the
theory have to be made without the assumption of externally given observers
\cite{ICQT} and initial/boundary conditions, thus excluding the
applicability of existing approaches such as Schwinger's action principle and
Feynman's path integral. Nevertheless, the two mentioned approaches motivate
us to suppose, as a basic postulate (\textquotedblleft the state-dynamics
postulate\textquotedblright) of the ICQFT, that the Universe is self-created
from no spacetime and no matter with the least action ($\hbar=1$):
\begin{align}
&  \left.  \left\vert \mathcal{B};A...,\psi...\right\rangle
=e^{iS(\mathcal{\hat{B}};\hat{A}...,\hat{\psi}...)}\left\vert \emptyset
\right\rangle \right. \nonumber\\
&  \left.  \delta S(\mathcal{\hat{B}};\hat{A}...,\hat{\psi}...)\left\vert
\mathcal{B};A...,\psi...\right\rangle =0 \right.
\label{postuD}%
\end{align}
Here $\left\vert \emptyset\right\rangle \equiv\left\vert \emptyset
_{\mathrm{G}}\right\rangle \otimes\left\vert \emptyset_{\mathrm{M}%
}\right\rangle $ is the common empty state of matter (the empty-matter
$\left\vert \emptyset_{\mathrm{M}}\right\rangle $) and geometry (the
empty-geometry state $\left\vert \emptyset_{\mathrm{G}}\right\rangle $ in loop
quantum gravity \cite{Rovelli-book}). Note that the requirement of the least
action gives the constraint conditions and the equations of motion as usual,
but with an interesting new feature, namely, the kinematics and dynamics of the
theory is indivisible. To see the feature, note that the dynamical law and
states always appear jointly in the postulate (Eq.~\ref{postuD}). This is in
sharp contrast to the tradition where the dynamical law [i.e., $\delta
S(\mathcal{\hat{B}};\hat{A}...,\hat{\psi}...)=0$] and states are given
separately. Thus, quantum entanglement dynamics of the trinary fields under
study unifies the dynamical law and states, a feature required by the
information-complete trinary description.

To summarize, the information-completeness principle puts a profound
restriction on what are physical systems and how to describe the physical systems.
Quantum relationalism stated above gives precisely an information-interpretation
of the gauge invariance in conventional QFT under local
Lorentz transformations, local gauge transformations, and diffeomorphism
\cite{Rovelli-book}. Namely, given the conventional gauge invariance, complete
information of the trinary fields is invariant. However, it seems that the
information completeness puts stronger restrictions on our field-theoretical
description than the usual gauge invariance; for instance, either $\hat{\psi}$
or $\hat{A}$ alone is information-incomplete field in the ICQFT. To be more
clear on this point, let us recall that a harmonic oscillator in classical
mechanics can have any continuous positive energy. But quantum mechanics
restricts its energy to be discrete values such that the oscillator's
classical state space that is previously physical according to classical
principles is now severely constrained by quantum principles. Here the
situation is completely the same. The information-completeness principle, or
put differently, the kinematics-dynamics indivisibility of the theory, 
severely restricts the allowed state space of the quantized fields such that
the originally physical states in conventional QFT become
unphysical in our new description. Below let us show how the ICQFT leads to a conceptually
compelling picture of unifying gravity and matter.

\section{Quantum state of gravity and matter}

The classical Einstein field equation reads \cite{Rovelli-book}:
\begin{equation}
R_{\mu}^{I}-\frac{1}{2}(R-2\Lambda)e_{\mu}^{I}=8\pi GT_{\mu}^{I} \label{EE}%
\end{equation}
where $R_{\mu}^{I}$ ($R$) is the Ricci tensor (scalar), $T_{\mu}^{I}$ the
energy-momentum tensor of matter; $\Lambda$ ($G$) represents the cosmological
(Newton) constant. In classical domains, Einstein's equation is extremely
successful. But quantum mechanically, it looks problematic as one has to
quantize these fields therein. As argued by Thiemann \cite{Thiemann}, in
quantum gravity theory $T_{\mu}^{I}$ should be quantized to be a field
operator $\hat{T}_{\mu}^{I}(\hat{e},\hat{\omega})$ in the Hilbert space of
both spacetime and matter. However, the problem (called hereafter as the
\textquotedblleft Hilbert-space inconsistency\textquotedblright, which also
applies to usual interacting quantized fields) still exists: Both sides of
Einstein's equation belong to different Hilbert spaces as the left are purely
operators for spacetime geometry; generally there is no way of equating them.
To remedy the inconsistency, one could of course act both sides of the
quantized Einstein equation upon a joint state of spacetime and matter such
that the equality for field operators is mapped into the equality for
classical field variables as in Eq.~\ref{EE}; see also the second
line of Eq.~\ref{postuD}. Then another problem arises as to what the joint
state of spacetime and matter is. As we will show below, the ICQFT provides a
concrete way to find the appropriate joint state of spacetime and matter to
\textquotedblleft glue\textquotedblright\ the two pieces of Einstein's equation.

Conceptual inconsistencies and difficulties of formulating a concrete quantum
gravity theory motivates the idea that Einstein's equation is merely an
effective spacetime theory \cite{Jacobson,Verlinde}; it cannot be quantized at
all in a way that we quantize matter fields. Thanks to the development of loop
quantum gravity, some conceptual inconsistencies and difficulties of the
quantum gravity theory have been overcome.

For the trinary fields under study, we need to specify a conservative quantity
(commutative with the gravitational Hamiltonian and with the matter
Hamiltonian) as a programming observable \cite{ICQT}. As we noted in the
context of the ICQT, the choice of the programming observable is relative so
that we could use gravity to programme matter, or vice versa. As the matter field couples to gravity via its
energy-momentum tensor (keeping in mind that conservation of the
energy-momentum tensor in curved spacetime is a subtle issue), it seems to be
natural to programme gravity by matter. However, as gravity couples
universally to all matter species, it is conceptually compelling to programme
matter by gravity. In this case, the Hamiltonian of the trinary fields can
then be formally written as:
\begin{align}
H_{\mathrm{G+M}}=\int d^{3}x\mathcal{H}_{\mathrm{M}\text{+}\mathrm{G}%
}(\hat{\psi},\hat{A};\mathcal{\hat{B}}_{a}^{i},\mathcal{\hat{E}}_{b}%
^{j})\nonumber\\
=\int d^{3}x\mathcal{H}_{\mathrm{G}}(\mathcal{\hat{B}}_{a}^{i}%
,\mathcal{\hat{E}}_{b}^{j})\nonumber\\
+\int d^{3}x\mathcal{H}_{\mathrm{M}\text{-}%
\mathrm{G}}(\hat{\psi},\hat{A};\mathcal{\hat{B}}_{a}^{i},\mathcal{\hat{E}}%
_{b}^{j})\nonumber\\
=H_{\mathrm{G}}(\mathcal{\hat{B}}_{a}^{i},\mathcal{\hat{E}}_{b}%
^{j})+H_{\mathrm{M}\text{-}\mathrm{G}}(\hat{\psi},\hat{A};\mathcal{\hat{B}%
}_{a}^{i},\mathcal{\hat{E}}_{b}^{j}) \label{H-GM}%
\end{align}
where $\mathcal{H}$ denotes the Hamiltonian density ($\mathcal{H}_{\mathrm{G}%
}$ for gravity, $\mathcal{H}_{\mathrm{M}\text{+}\mathrm{G}}$ for gravity and
matter, and $\mathcal{H}_{\mathrm{M}\text{-}\mathrm{G}}$ for the
gravity-matter coupling); $\bar{\psi}$ and momentum conjugate to $\hat{A}$ are
all omitted for notation simplicity.

Although the matter-field sector of the problem is less developed
\cite{Rovelli-book,Thiemann-reg,Chiou}, the gravity sector is well established
within loop quantum gravity so that, with the input of the
information-completeness principle, we can write the dual spacetime-matter
entangled state, resulting from the state-dynamics postulate, in the standard
Schmidt form as:
\begin{equation}
\left\vert \mathcal{B},(\psi,A)\right\rangle =\sum_{s}\mathrm{S}%
_{\mathrm{G+M}}[s]\left\vert \mathcal{B},s\right\rangle \otimes
\text{$\left\vert (\psi,A),s\right\rangle $} \label{Ent-GM}%
\end{equation}
Here $\mathrm{S}_{\mathrm{G+M}}[s]$ ($>0$) denotes the Schmidt coefficients
and is determined by dynamics of the trinary system; \{$\left\vert
\mathcal{B},s\right\rangle$\} (named as the \textquotedblleft programming
basis\textquotedblright\ in the ICQT) and \{$\left\vert (\psi,\notag\right. \\ \left.A),s\right\rangle
$\} span orthogonal bases for the Hilbert spaces of spacetime and matter,
respectively. Without loss of generality we assume $s$ to be discrete. In loop
quantum gravity we do have the spin-network basis of discrete spectra
\cite{Rovelli-book,Chiou}. With respect to this specific decomposition, the
programmed entangled state $\left\vert (\psi,A),s\right\rangle $ for the Dirac
field and the electromagnetic field can be likewise decomposed as:
\begin{equation}
\text{$\left\vert (\psi,A),s\right\rangle $}=\sum_{\ell}\mathrm{S}%
_{\mathrm{M}\left\vert \mathrm{G}\right.  }[\ell,s]\left\vert A,\ell
,s\right\rangle \otimes\left\vert \psi,\ell,s\right\rangle  \label{Ent-DE}%
\end{equation}
which encodes complete information about $\hat{\psi}$ and $\hat{A}_{\mu}$ as
programmed by $\left\vert \mathcal{B},s\right\rangle $. \{$\left\vert
\psi,\ell,s\right\rangle $\} and \{$\left\vert A,\ell,s\right\rangle $\} are
two orthogonal bases for the Hilbert spaces of the Dirac field and the
electromagnetic field.

Let us consider the physical significance of dual entanglement in
Eqs.~\ref{Ent-GM}-\ref{Ent-DE}. Dual entanglement in Eqs.~\ref{Ent-GM}%
-\ref{Ent-DE} already implies the existence of discrete orthonormal bases
\{$\left\vert \mathcal{B},s\right\rangle $\}, \{$\left\vert (\psi
,A),s\right\rangle $\}, \{$\left\vert \psi,\ell,s\right\rangle $\}, and
\{$\left\vert A,\ell,s\right\rangle $\}, as a direct consequence of our
information-complete description. The information-completeness in turn implies
that all these states [$\left\vert \mathcal{B},s\right\rangle $, $\left\vert
(\psi,A),s\right\rangle $,\ $\left\vert \psi,\ell,s\right\rangle $,
and\ $\left\vert A,\ell,s\right\rangle $] involved in the dual Schmidt forms
have to be physical states in their corresponding state spaces, eliminating
any gauge arbitrariness, of the trinary system as they encode all the relevant
information of direct \textit{physical} significance (namely, physical
predictions of the theory).

One of the most important results achieved by loop quantum gravity is the
identification of the \textquotedblleft spin-network\textquotedblright\ states
as the Hilbert space of quantized gravity. These states are
diffeomorphism-invariant, form a discrete orthonormal basis, and support
discrete spacetime geometry
\cite{Thiemann,quanA1,quanA2,quanA3,loop25,Rovelli-book,Thiemann-reg,Chiou}.
Now an important problem arises: What is the relation between the spin-network
states and $\left\vert \mathcal{B},s\right\rangle $ in Eq.~\ref{Ent-GM}?
Below we will give first of all a rigorous (but formal) formulation of the
ICQFT. Then we make a conjecture, in which the spin-network states span the
programming basis.

\section{Dual dynamics and solution to problem of time}

As constraint physical systems, the trinary fields satisfy the Gauss, the
diffeomorphism, and the Hamiltonian constraints which must annihilate the
physical state \cite{Rovelli-book,Thiemann-reg,Chiou}, e.g.:
\begin{equation}
H_{\mathrm{G+M}}\left\vert \mathcal{B},(\psi,A)\right\rangle
=0 \label{Hconstr}%
\end{equation}
for the Hamiltonian constraint. As a result of Eq.~\ref{Hconstr}, the whole
system seems to have no time evolution, a fact known as the \textquotedblleft
problem of time\textquotedblright\ \cite{Rovelli-book,Chiou,Kiefer} in quantum
gravity and quantum cosmology. As we noted above, for our purpose we need a
programming observable, commutative with the gravitational Hamiltonian
$H_{\mathrm{G}}(\mathcal{\hat{B}}_{a}^{i},\mathcal{\hat{E}}_{b}^{j})$ and the
total Hamiltonian $H_{\mathrm{G+M}}$. Now let us show how we can obtain the observable.

The spacetime-matter state $\left\vert \mathcal{B},(\psi,A)\right\rangle $ can
of course be expanded in any orthogonal bases for the Hilbert spaces of
gravity and matter. We thus can freely choose a basis such that the reduced
density operator for gravity has only positive diagonal elements, namely:
\begin{align}
\rho_{\mathrm{G}}^{\{s\}} &  =\mathrm{tr}_{\mathrm{M}}[\left\vert
\mathcal{B},(\psi,A)\right\rangle \left\langle \mathcal{B},(\psi,A)\right\vert
]\nonumber\\
&  =\sum_{s}\mathrm{S}_{\mathrm{G+M}}^{2}[s]\left\vert \mathcal{B}%
,s\right\rangle \left\langle \mathcal{B},s\right\vert \label{doG}%
\end{align}
where $\mathrm{tr}_{\mathrm{M}}$ means trace over the matter state space. Note
that this diagonal form of $\rho_{\mathrm{G}}^{\{s\}}$ can always be achieved
by a unitary transformation upon the Hilbert space \textit{merely for
gravity}. We can also rewrite Eq.~\ref{doG} as:
\begin{equation}
\rho_{\mathrm{G}}^{\{s\}}\equiv\frac{e^{-\mathcal{\hat{I}}_{\mathrm{Eu}}}%
}{\mathrm{tr}(e^{-\mathcal{\hat{I}}_{\mathrm{Eu}}})}\equiv\frac{e^{-\beta
\hat{\Xi}}}{\mathrm{tr}(e^{-\beta\hat{\Xi}})} \label{eH}%
\end{equation}
where $\beta=1/(\kappa_{B}T)$ with $\kappa_{B}$\ being the Boltzmann constant
and $T$ having dimension of temperature; a Hermitian operator $\hat{\Xi}$,
which is positive semidefinite, is named as the \textquotedblleft entanglement
Hamiltonian\textquotedblright\ in other contexts \cite{eH-Li,eH-1Dfermi}.
Hereafter we call $\mathcal{\hat{I}}_{\mathrm{Eu}}$ the \textit{Euclidean
entanglement action}, whose spectrum contains complete information of
$\rho_{\mathrm{G}}^{\{s\}}$ and thus, spacetime-matter (but not matter-matter)
entanglement. We conjecture that $\mathcal{\hat{I}}_{\mathrm{Eu}}$ might be
given by certain Euclidean action of gravity. We leave this conjecture for
future work; for the Euclidean action of gravity, see, e.g.,
Refs.~\cite{EuA-Hawking,EuA-Pad}.

The information-completeness within a trinary description demands that:
\begin{align}
H_{\mathrm{G+M}}  &  =H_{\mathrm{G}}+H_{\mathrm{M}\text{-}\mathrm{G}%
}\nonumber\\
H_{\mathrm{M}\text{-}\mathrm{G}}  &  =\sum_{s}\left\vert \mathcal{B}%
,s\right\rangle \left\langle \mathcal{B},s\right\vert \otimes H_{\mathrm{M}%
\left\vert \mathrm{G}\right.  }(\hat{\psi},\hat{A};s) \label{totalH}%
\end{align}
where gravity and matter are coupled with $H_{\mathrm{M}\text{-}\mathrm{G}}$
of a factorizable form, and $H_{\mathrm{M}\left\vert \mathrm{G}\right.  }%
(\hat{\psi},\hat{A};s)$ is the matter Hamiltonian conditional on (namely,
programmed by) the gravity state $\left\vert \mathcal{B},s\right\rangle $. The
overall evolution of the gravity-matter system is given in Heisenberg's
picture by:
\begin{align}
\left\vert \mathcal{B},(\psi,A)\right\rangle  &  =\hat{U}_{\mathrm{G+M}%
}(t)\left\vert \emptyset\right\rangle \nonumber\\
\hat{U}_{\mathrm{G+M}}(t)  &  =\sum_{s}\left\vert \mathcal{B},s\right\rangle
\left\langle \mathcal{B},s\right\vert \hat{U}_{\mathrm{G}}(t)\otimes\hat
{U}_{\mathrm{M}\left\vert \mathrm{G}\right.  }(s,t) \label{totalU}%
\end{align}
The evolution operator $\hat{U}_{\mathrm{G+M}}$ also has a factorizable
structure and is determined by:
\begin{align}
i\frac{\partial}{\partial t}\hat{U}_{\mathrm{G}}(t)  &  =H_{\mathrm{G}}\hat
{U}_{\mathrm{G}}(t)\nonumber\\
i\frac{\partial}{\partial t}\hat{U}_{\mathrm{M}\left\vert \mathrm{G}\right.
}(s,t)  &  =H_{\mathrm{M}\left\vert \mathrm{G}\right.  }\hat{U}_{\mathrm{M}%
\left\vert \mathrm{G}\right.  }(s,t) \label{pairU}%
\end{align}
As we have chosen a particular basis for gravity as in Eq.~\ref{doG}, the
basic property of the Schmidt decomposition \cite{Preskill} leads to that all
$\left\vert (\psi,A),s;t\right\rangle \equiv\hat{U}_{\mathrm{M}\left\vert
\mathrm{G}\right.  }(s,t)\left\vert \emptyset_{\mathrm{M}}\right\rangle $ form
an orthonormal basis and
\begin{equation}
\hat{U}_{\mathrm{G}}(t)\left\vert \emptyset_{\mathrm{G}}\right\rangle
=\sum_{s}\mathrm{S}_{\mathrm{G+M}}[s,t]\left\vert \mathcal{B},s\right\rangle 
\label{Pstate}%
\end{equation}

The dynamical evolutions in Eqs.~\ref{totalU} and \ref{pairU} take the
desired form as in the ICQT. They are mutually defined for spacetime and
matter as expected. A similar dynamics can be obtained for the matter fermion
field and the gauge field. The dual dynamical evolution always results in
correct dual entanglement, in which \textit{all constituent states are ensured
to be physical}. Now it is ready to see that dual dynamics is a robust feature
of our information-complete trinary description. As can be seen from
Eq.~\ref{Hconstr}, using Eqs.~\ref{totalU} and \ref{pairU} yields:
\begin{equation}
\frac{\partial}{\partial t}\hat{U}_{\mathrm{G+M}}(t)=0\label{timeless}%
\end{equation}
namely, the whole system (spacetime+matter) cannot have a dynamical evolution,
indeed. Yet, both spacetime and matter have their own dynamical evolutions,
which are \textquotedblleft glued\textquotedblright\ by spacetime-matter
entanglement. Thus, the problem of time, remaining as one of the conceptual
obstacles for a consistent quantum gravity, disappears in our formalism.

Equation~\ref{Pstate} and the factorizable form of $\hat{U}_{\mathrm{G+M}%
}(t)$ have a physically appealing interpretation as follows. $\hat
{U}_{\mathrm{G}}(t)$, exactly like a quantum gate (the \textquotedblleft
gravity gate\textquotedblright), prepares the gravity state $\sum
_{s}\mathrm{S}_{\mathrm{G+M}}[s,t]\left\vert \mathcal{B},s\right\rangle $ as
the controlling state from $\left\vert \emptyset_{\mathrm{G}%
}\right\rangle $. Then the \textrm{controlled}-$\hat{U}_{\mathrm{M}\left\vert
\mathrm{G}\right.  }$ operation\ (the \textquotedblleft gravity-matter
gate\textquotedblright) $\sum_{s}\left\vert \mathcal{B},s\right\rangle
\left\langle \mathcal{B},s\right\vert \otimes\hat{U}_{\mathrm{M}\left\vert
\mathrm{G}\right.  }(s,t)$\ creates the gravity-matter entangled state
$\left\vert \mathcal{B},(\psi,A)\right\rangle $. Meanwhile, $\hat
{U}_{\mathrm{M}\left\vert \mathrm{G}\right.  }(s,t)$ completely determines the
entanglement between matter fermions and their gauge fields; the number of
independent $\hat{U}_{\mathrm{M}\left\vert \mathrm{G}\right.  }(s,t)$ equals
the Schmidt number of $\left\vert \mathcal{B},(\psi,A)\right\rangle $. In this
quantum-gate interpretation of $\hat{U}_{\mathrm{G+M}}(t)=e^{iS(\mathcal{\hat
{B}};\hat{A}...,\hat{\psi}...)}$, the state-dynamics postulate $\delta
S(\mathcal{\hat{B}};\hat{A}...,\hat{\psi}...)\left\vert \mathcal{B}%
;A...,\psi...\right\rangle =0$ might be equivalent to maximizing entanglement
(information) with the least \textquotedblleft gate action\textquotedblright.

The central point of the ICQFT is that \textit{the physical predictions are
dual entanglement }$\left\vert \mathcal{B},(\psi,A)\right\rangle $\textit{ in
the Schmidt form, which encodes complete information on how gravity and matter
are entangled and how matter fermions and their gauge fields are entangled as
programmed by gravity}. In particular, the reduced density operator
$\rho_{\mathrm{G}}^{\{s\}}$ in Eq.~\ref{doG} for gravity (similarly for
matter) is the physical predictions and thus must be also a physical
observable (i.e., the \textquotedblleft complete observable\textquotedblright%
,\ also known as Dirac's observable) of the theory. This in turn implies that
\{$\left\vert \mathcal{B},s\right\rangle $\} is the programming basis and
$\rho_{\mathrm{G}}^{\{s\}}$ (or, $\mathcal{\hat{I}}_{\mathrm{Eu}}$ and
$\hat{\Xi}$), the programming observable, must commute with all the
constraints of the theory, e.g.:
\begin{equation}
\lbrack\rho_{\mathrm{G}}^{\{s\}},H_{\mathrm{G+M}}]=[\rho_{\mathrm{G}}%
^{\{s\}},H_{\mathrm{G}}+H_{\mathrm{M}\text{-}\mathrm{G}}]=0\label{rogm0}%
\end{equation}
for the Hamiltonian constraint. Here and hereafter all commutators are
understood to act upon $\left\vert \mathcal{B},(\psi,A)\right\rangle $ because
of the state-dynamics postulate (Eq.~\ref{postuD}). As can be easily checked,
$[\rho_{\mathrm{G}}^{\{s\}},H_{\mathrm{M}\text{-}\mathrm{G}}]=0$, we have
\begin{equation}
\lbrack\rho_{\mathrm{G}}^{\{s\}},H_{\mathrm{G}}]=0 \label{roghg}%
\end{equation}
as a result of Eq.~\ref{rogm0}. Thus, under the chosen particular basis for
gravity, the gravity Hamiltonian $H_{\mathrm{G}}$ is itself a physical
observable. Interestingly, Eqs.~\ref{rogm0} and \ref{roghg} indicate that
$\rho_{\mathrm{G}}^{\{s\}}$ (as well as $H_{\mathrm{G}}$) is also a quantum
nondemolition observable (For detailed discussions on quantum nondemolition
observables, see Ref.~\cite{QND-rmp}).

The fact that $\rho_{\mathrm{G}}^{\{s\}}$ and $H_{\mathrm{G}}$ are physical
observables of the theory ensures the consistency of the above considerations
on dual dynamics. As a result of Eq.~\ref{roghg}, $\left\vert \mathcal{B}%
,s\right\rangle $ is an eigenstate of $H_{\mathrm{G}}$ with eigenvalue
$E_{\mathrm{G}}(s)$; the ordering of $\hat{U}_{\mathrm{G}}(t)$ and $\left\vert
\mathcal{B},s\right\rangle \left\langle \mathcal{B},s\right\vert $ in $\hat
{U}_{\mathrm{G+M}}(t)$ (see Eq.~\ref{totalU}) is thus not important.
Meanwhile, it is easy to prove that in Eq.~\ref{Pstate}:
\begin{equation}
\mathrm{S}_{\mathrm{G+M}}[s,t]=\mathrm{S}_{\mathrm{G+M}}[s]e^{-itE_{\mathrm{G}%
}(s)} \label{Smt}%
\end{equation}
such that the time-dependence of $\mathrm{S}_{\mathrm{G+M}}[s,t]$ is solely
from $e^{-itE_{\mathrm{G}}(s)}$.

To illustrate the dynamics of the trinary fields further, we can also work in
Schr\"{o}dinger's picture. To this end, note first that it is meaningless to
consider the time evolution of the whole system in $\left\vert \mathcal{B}%
,(\psi,A)\right\rangle $, which is timeless and encodes complete physical
information on the whole spacetime (of course the whole time) and all matter
contents (if we include all matter Hamiltonians in $H_{\mathrm{G+M}}$).
However, it is still meaningful to consider the time evolution of an
individual constituent state $\left\vert \mathcal{B},s;t\right\rangle
\otimes\left\vert (\psi,A),s;t\right\rangle $ of $\left\vert \mathcal{B}%
,(\psi,A)\right\rangle =\sum_{s}\mathrm{S}_{\mathrm{G+M}}[s]\left\vert
\mathcal{B},s;t\right\rangle \otimes\left\vert (\psi,A),s;t\right\rangle $, in
which the time-dependence of $\left\vert \mathcal{B},s;t\right\rangle $ and
$\left\vert (\psi,A),s;t\right\rangle $ is explicitly shown. Note that, as a
result of Eq.~\ref{Hconstr}:
\begin{equation}
(H_{\mathrm{G}}+H_{\mathrm{M}\text{-}\mathrm{G}})\left\vert \mathcal{B}%
,(\psi,A)\right\rangle =i\frac{\partial}{\partial t}\left\vert \mathcal{B}%
,(\psi,A)\right\rangle =0 \label{HgHmg0}%
\end{equation}
such that:
\begin{align}
&  \sum_{s}\mathrm{S}_{\mathrm{G+M}}[s][H_{\mathrm{G}}\left\vert
\mathcal{B},s;t\right\rangle \otimes\left\vert (\psi,A),s;t\right\rangle
\nonumber\\
&  \left.  +\left\vert \mathcal{B},s;t\right\rangle \otimes H_{\mathrm{M}%
\left\vert \mathrm{G}\right.  }(\hat{\psi},\hat{A};s)\left\vert (\psi
,A),s;t\right\rangle ]=0\right. \label{sequal}%
\end{align}

Let us define $\left\vert (\psi,A),s;t\right]  \equiv\left\langle
\mathcal{B},s;t\right.  \left\vert \mathcal{B},(\psi,A)\right\rangle
=\mathrm{S}_{\mathrm{G+M}}[s]\left\vert (\psi,A),s;t\right\rangle $. Note that
$\left\vert (\psi,A),s;t\right]  $ is unnormalized and its inner product
$\left[  (\psi,A),s;t\right.  \left\vert (\psi,A),s;t\right]  =\mathrm{S}%
_{\mathrm{G+M}}^{2}[s]$ represents the probability of finding the whole system
in $\left\vert \mathcal{B},s;t\right\rangle $. Now if we require that:
\begin{equation}
i\frac{\partial}{\partial t}\left\vert \mathcal{B},s;t\right\rangle
=H_{\mathrm{G}}(\mathcal{\hat{B}}_{a}^{i},\mathcal{\hat{E}}_{b}^{j})\left\vert
\mathcal{B},s;t\right\rangle  \label{SchG}%
\end{equation}
then using Eqs.~\ref{HgHmg0} and \ref{sequal} we have $i\frac{\partial
}{\partial t}\left\vert (\psi,A),s;t\right]  =\left\langle \mathcal{B}%
,s;t\right\vert H_{\mathrm{M}\text{-}\mathrm{G}}\left\vert \mathcal{B}%
,(\psi,A)\right\rangle $. If $\mathrm{S}_{\mathrm{G+M}}[s]$ is
time-independent as it must be in Schr\"{o}dinger's picture, we have at once:
\begin{equation}
i\frac{\partial}{\partial t}\left\vert (\psi,A),s;t\right\rangle
=H_{\mathrm{M}\left\vert \mathrm{G}\right.  }(\hat{\psi},\hat{A}%
,s;t)\left\vert (\psi,A),s;t\right\rangle  \label{SchMG}%
\end{equation}
in which the orthogonality of \{$\left\vert \mathcal{B},s;t\right\rangle $\}
and \{$\left\vert (\psi,A),s;t\right\rangle $\} is used.

The dynamical equations as given in Eqs.~\ref{SchG} and \ref{SchMG} solve the
problem of time in Schr\"{o}dinger's picture. The solution resembles the
Page-Wootters mechanism \cite{Page-Wootters} and in particular its very recent
version \cite{QTime}. However, in the context of quantum mechanics it is hard
to associate time with a quantum degree of freedom. Fortunately, in quantum
gravity we do have the desired quantum degree of freedom as spacetime itself
is quantized.

Another problem related to the problem of time is how to reconcile the
apparent macroscopic irreversibility (e.g., the second law of thermodynamics)
with the time-symmetry of microscopic laws. This is known as the paradox of
time's arrow that has puzzled physicists at least since Boltzmann. Here,
rather than reviewing the history of this long-standing problem, we give a simple 
argument showing that our theory might have an arrow of
time. In QFT, the time symmetry is embodied by the invariance
under a time-reversion operator $\hat{T}$ ($\hat{T}_{\mathrm{G}}$ for gravity
and $\hat{T}_{\mathrm{M}}$ for matter), which can be defined for every
quantized field. A basic property of entanglement is that it does not decrease under any local unitary
transformations (such as charge conjugation $\hat{C}$ and space inversion $\hat{P}$). 
If we could assume that entanglement also does not decrease under any local anti-unitary
transformations like time inversion $\hat{T}$---an unproven statement to the best of the author's knowledge, 
then in our case, even if the theory has time-symmetry in the
usual sense, dual entanglement in the time-inversion state $\hat
{T}_{\mathrm{G}}\hat{T}_{\mathrm{M}}\left\vert \mathcal{B},(\psi
,A)\right\rangle $ never decreases under time inversion (e.g., $\hat{T}_{\mathrm{G}}$ and $\hat{T}_{\mathrm{M}}$) upon
the Hilbert space of each field. If this is indeed the case, our theory could have an
\textit{entanglement-induced arrow of time} \cite{ICQT} and allow
time-asymmetry at the most fundamental level; for a further discussion supporting the above argument on this
issue, see below, Section {\bf 9}. The $\mathrm{CPT}$ symmetry for spin-foam fermions was
discussed in Ref.~\cite{Spin-Fermi}; full consideration of the $\mathrm{CPT}$
symmetry in our theory will be given in future.

\section{Conceptual aspects of information-complete description}

Equations \ref{SchG} and \ref{SchMG} imply that Einstein's equation in
quantum domain is separated into two pieces, one purely for spacetime and
another for matter programmed by spacetime. This eliminates the Hilbert-space
inconsistency of Einstein's equation. It should be emphasized that in our
picture, gravity plays a unique role as the programming system, whose Hilbert
space supports information-complete field operators \cite{ICQT}. By contrast,
either the fermion field or its gauge field alone is information-incomplete;
only jointly they are information-complete physical entity. Thus, quantizing
gravity alone is meaningful within current loop quantum gravity of remarkable
success. On the other hand, we must put the information-completeness and the
matter contents into a trinary description to complete a consistent quantum
theory of gravity coupled with matter, thus paving the way to consistently
quantize the matter sector as well. However, the matter sector within
quantized spacetime is currently not well understood and progress has been
made steadily \cite{Thiemann-reg,Rovelli-eigen,cano-Ferimi,Spin-Fermi}. In
this regard, it is reasonable to expect that the ICQFT will play a role in
further development on quantum gravity, especially on quantization of the
matter sector.

One should notice a subtle issue in the above considerations. When we use the
evolution $\left\vert \mathcal{B},(\psi,A)\right\rangle =\hat{U}%
_{\mathrm{G+M}}(t)\left\vert \emptyset\right\rangle $ with a factorizable
$\hat{U}_{\mathrm{G+M}}(t)$ (see Eq.~\ref{totalU}), the matter states
$\left\vert (\psi,A),s;t\right\rangle =\hat{U}_{\mathrm{M}\left\vert
\mathrm{G}\right.  }(s,t)\left\vert \emptyset_{\mathrm{M}}\right\rangle $ must
span an orthonormal basis such that they are physical as well. This statement,
together with Eq.~\ref{pairU} and the existence of the programming basis
\{$\left\vert \mathcal{B},s\right\rangle $\} for gravity, can be regarded as
the \textit{definition of physical Hamiltonians for matter and gravity}. The
usual Hamiltonians for matter and gravity are obtained from the gravity-matter
action under the requirements of the invariance under local Lorentz
transformations, local gauge transformations, and diffeomorphism. Are they
identical to the physical Hamiltonians for matter and gravity as required by
our theory? If the answer to this open question is \textquotedblleft
no\textquotedblright, then it is ready to see that our information-complete
trinary description puts stronger and more restrictions on quantum fields than
the usual formalism, as we pointed out above. The restrictions, which modify
the usual Hamiltonians, stem from and are enforced by dual entanglement.

Now it is time for some remarks about the general feature of the ICQFT, which
is a field-theoretic generalization of a new quantum formalism \cite{ICQT}
developed recently. As the information-complete trinary description, the ICQFT
does not require the measurement postulate and shares dual dynamics and dual
entanglement structure of the trinary fields; kinematics and dynamics is
indivisible, too: While all dynamical information is completely encoded in
dual entanglement of spacetime and matter, kinematics about states and
observables for an individual field is either meaningless or
information-incomplete; only states and observables involved in dual
entanglement (joint properties of the trinary fields) are of dynamical and
physical significance. In this way, a huge number of unphysical degrees of
freedom, while appearing in conventional QFT, is eliminated.
Moreover, the ICQFT has a uniquely determined \textquotedblleft initial
condition\textquotedblright, namely, $\left\vert \emptyset\right\rangle $,
which is a physical state and means \textit{absolute nothing and nowhere}, in
sharp contrast to the concept of vacuum in conventional QFT.
As spacetime and matter are mutually defined via spacetime-matter entanglement
in the ICQFT, $\left\vert \emptyset\right\rangle $ is a state of no matter and
no spacetime and, particularly, does not correspond to a flat spacetime, which
is simply empty and meaningless in our theory as there is no matter to define it.

In the information-complete trinary description under study, all matter fields and
even spacetime are quantized, and as such there is simply no any room for classical entities
such as meaurement apparatus and observers. Then in such a description, the theory has to
consider a self-defining/explaining quantum world without appealing to any physical entity
beyond the theoretic structure itself. The quantum formalism given here has the 
feature as stated. Let us recall that, even in loop quantum gravity 
(as well as in superstring theory, of course), 
the interpretational and conceptual problems of quantum foundations remain 
there—They root deeper than quantization of spacetime and call for a major conceptual 
step for their resolution. In other words, current quantum theory is itself an 
unfinished revolution, not simply because spacetime is not quantized, but 
rather because its own formulation calls for a radical change. 

\section{Spin-network states as programming states}

In the formulation of the ICQFT given above, it is of course advantageous to
have the explicit form of the programming basis \{$\left\vert \mathcal{B},s\right\rangle $\}. 
Here we give argument which supports the spin-network
states spanning the programming basis. Without the input of known results for
loop quantum gravity, the follow-up considerations would be too formal and of
less predictive power.

For an abstract graph $\Gamma$ (with nodes labeled by $n$ and oriented links
labeled by $l$) in three-dimensional region $\mathcal{R}$ with two-dimensional
surface $\digamma$ embedded, a spin-network state $\left\vert \Gamma
,j_{l},i_{n}\right\rangle $, where $j_{l}$ is an irreducible $j$
representation of $SU(2)$ for each link $l$ and $i_{n}$ the $SU(2)$
intertwiner for each node $n$, is the common eigenstates
\cite{Rovelli-book,Chiou} of the area operator $\mathbf{\hat{A}}(\digamma)$
[with eigenvalue $\mathbf{A}(j_{l})$ for the link $l$] and the volume operator
$\mathbf{\hat{V}}(\mathcal{R})$ [with eigenvalue $\mathbf{V}(i_{n})$ for the
node $n$]. Then, $\left\vert \Gamma,\mathbf{J},\mathbf{I}\right\rangle
=\left\vert \Gamma,j_{1}...j_{L},i_{1}...i_{N}\right\rangle $ represents a
spin-network state for $N$ quanta of volume, separated from each other by the
adjacent surfaces of $L$ quanta of area. The spin-network states, once defined
in a diffeomorphism invariant way, span an orthonormal basis ($\left\langle
\Gamma,\mathbf{J}^{\prime},\mathbf{I}^{\prime}\right\vert \left.
\Gamma,\mathbf{J},\mathbf{I}\right\rangle =\delta_{\Gamma^{\prime}\Gamma
}\delta_{\mathbf{J}^{\prime}\mathbf{J}}\delta_{\mathbf{I}^{\prime}\mathbf{I}}%
$) for the physical Hilbert space of quantized gravity.

On the other hand, quantum states of gravity coupled with matter in loop
quantum gravity read \cite{Rovelli-book,Chiou}:
\begin{equation}
\left\vert \Gamma,j_{l},i_{n}\right\rangle \otimes\left\vert k_{l},F_{n}%
,w_{n}\right\rangle  \label{corr}%
\end{equation}
where $k_{l}$ is the electric flux across the surface $l$ and $F_{n}$ ($w_{n}%
$) represents the number of fermions (field strength) at node $n$; the Higgs
field is not included and will be considered elsewhere \cite{gGUT}; here the
number $\bar{F}_{n}$ of anti-fermions at node $n$ is not included for notation
simplicity. These states show explicitly the correlations between the
spin-network states $\left\vert \Gamma,j_{l},i_{n}\right\rangle $ and the
matter states $\left\vert k_{l},F_{n},w_{n}\right\rangle $, similarly to the
correlations between $\left\vert \mathcal{B},s\right\rangle $ and $\left\vert
(\psi,A),s\right\rangle $ in Eq.~\ref{Ent-GM}. Therefore, it is a natural
assumption to identify $\left\vert \mathcal{B},s\right\rangle $ with
$\left\vert \Gamma,j_{l},i_{n}\right\rangle $. Consequently, the spin-network
states $\left\vert \Gamma,j_{l},i_{n}\right\rangle $ are physical prediction,
and the geometry operators $\mathbf{\hat{A}}(\digamma)$ and $\mathbf{\hat{V}%
}(\mathcal{R})$\ defined in a diffeomorphism invariant manner
\cite{Thiemann,Chiou} are physical observables.

Then $H_{\mathrm{M}\text{-}\mathrm{G}}(\hat{\psi},\hat{A};\mathcal{\hat{B}%
}_{a}^{i},\mathcal{\hat{E}}_{b}^{j})$ in Eq.~\ref{H-GM} can be expanded in
terms of $\left\vert \Gamma,j_{l},i_{n}\right\rangle $ as:
\begin{equation}
H_{\mathrm{M}\text{-}\mathrm{G}}^{\{\Gamma\}}=\sum_{\substack{l\in\Gamma
\cap\digamma\\\Gamma,n\in\mathcal{R}}}\left\vert \Gamma,j_{l},i_{n}%
\right\rangle \left\langle \Gamma,j_{l},i_{n}\right\vert \otimes
H_{\mathrm{M}\left\vert \mathrm{G}\right.  }^{\Gamma(j_{l},i_{n})}(\hat{\psi
},\hat{A}) \label{prog-M}%
\end{equation}
where $H_{\mathrm{M}\left\vert \mathrm{G}\right.  }^{\Gamma(j_{l},i_{n})}%
(\hat{\psi},\hat{A})$ is the programmed matter Hamiltonian associated with
$\left\vert \Gamma,j_{l},i_{n}\right\rangle $ and $H_{\mathrm{M}%
\text{-}\mathrm{G}}^{\{\Gamma\}}\equiv H_{\mathrm{M}\text{-}\mathrm{G}}$. In
this way, the spin-network states, while defining the intrinsic geometry
\cite{Thiemann,Rovelli-book,Chiou}, are interpreted here to be correlated with
the states generated by the matter Hamiltonian to ensure that spacetime and
matter are mutually measured and entangled; $\left\vert \Gamma,j_{l}%
,i_{n}\right\rangle $ are then physical states (namely, the physical
predictions of the theory) that represent not merely geometry without matter contents.

As is already known in loop quantum gravity, the geometry operators
$\mathbf{\hat{A}}(\digamma)$ and $\mathbf{\hat{V}}(\mathcal{R})$\ are
\textquotedblleft partial observables\textquotedblright\ as named by Rovelli
\cite{Rovelli-book}. In a conceptually clear way, Thiemann \cite{Thiemann}
argued that the geometry operators become diffeomorphism invariant and thus
physical as soon as they couple with matter excitations. According to the
above-mentioned general feature of the ICQFT, spacetime-matter entanglement is
the deeper physics underlying Thiemann's argument. This is the very reason why
only those $\left\vert \Gamma,j_{l},i_{n}\right\rangle $ appearing in
spacetime-matter entanglement are physical predictions of our theory: Here the
geometry quanta are counted/measured only by matter excitations and there is
no counting if no matter excitations; in this sense, \textit{physical
geometry} is a joint property of spacetime \textit{and} matter. This leads to
a huge truncation of the spin-network states for gravitational Hilbert space
as imposed by the state-dynamics unification of our formalism, namely, the
indivisibility of kinematics and dynamics.

As $\left\vert \Gamma,j_{l},i_{n}\right\rangle $ are the common eigenstates of
$\mathbf{\hat{A}}(\digamma)$ and $\mathbf{\hat{V}}(\mathcal{R})$, one
obviously has:
\begin{equation}
\lbrack\mathbf{\hat{A}}(\digamma),H_{\mathrm{M}\text{-}\mathrm{G}}%
^{\{\Gamma\}}]=[\mathbf{\hat{V}}(\mathcal{R}),H_{\mathrm{M}\text{-}\mathrm{G}%
}^{\{\Gamma\}}]=0 \label{ahvh}%
\end{equation}
Meanwhile, $\mathbf{\hat{A}}(\digamma)$ and $\mathbf{\hat{V}}(\mathcal{R}%
)$\ as physical observables of the theory should commute with all the
constraints. In particular, we have:
\begin{equation}
\lbrack\mathbf{\hat{A}}(\digamma),H_{\mathrm{G}}+H_{\mathrm{M}\text{-}%
\mathrm{G}}^{\{\Gamma\}}]=[\mathbf{\hat{V}}(\mathcal{R}),H_{\mathrm{G}%
}+H_{\mathrm{M}\text{-}\mathrm{G}}^{\{\Gamma\}}]=0 \label{total}%
\end{equation}
implying:
\begin{equation}
\lbrack\mathbf{\hat{A}}(\digamma),H_{\mathrm{G}}]=[\mathbf{\hat{V}%
}(\mathcal{R}),H_{\mathrm{G}}]=0 \label{ahvh0}%
\end{equation}
as a result of Eq.~\ref{ahvh}. Equation~\ref{ahvh0} implies that
$\left\vert \Gamma,j_{l},i_{n}\right\rangle $ are also the eigenstates of the
Hamiltonian $H_{\mathrm{G}}$ for the gravity sector, and in particular:
\begin{equation}
\lbrack H_{\mathrm{G}},H_{\mathrm{M}\text{-}\mathrm{G}}^{\{\Gamma
\}}]=[H_{\mathrm{G}},H_{\mathrm{G+M}}]=0. \label{commu}%
\end{equation}

Similarly to the above general discussions on dynamics, we can also consider
the time evolution of an individual spin-network state (or any superposition
of a given set of spin-network states). In this case, the spin-network states
are not defined in spacetime; rather, they \textit{are} spacetime
\cite{Rovelli-book}. So we can include the explicit time- and
field-dependences for the spin-network states by $\left\vert \mathcal{B}%
,\Gamma,j_{l},i_{n};t\right\rangle $, as well as for the matter states by
$\left\vert (\psi,A),\Gamma,k_{l},F_{n},w_{n};t\right\rangle $ such that the
pair-equations as in (\ref{SchG}) and (\ref{SchMG}) can be obtained.
Meanwhile, in Heisenberg's picture Eq.~\ref{Ent-GM} is rewritten as:
\begin{align}
\left\vert \mathcal{B},(\psi,A)\right\rangle  &  =\sum_{\substack{l\in
\Gamma\cap\digamma\\\Gamma,n\in\mathcal{R}}}\mathrm{S}_{\mathrm{G+M}}%
[\Gamma,n,l;t]\left\vert \mathcal{B},\Gamma,j_{l},i_{n}\right\rangle
\nonumber\\
&  \otimes\left\vert (\psi,A),\Gamma,k_{l},F_{n},w_{n}\right\rangle
\label{Ent-t}%
\end{align}
which is generated by $\left\vert \mathcal{B},(\psi,A)\right\rangle =\hat
{U}_{\mathrm{G+M}}(t)\left\vert \emptyset\right\rangle $ with
\begin{align}
\hat{U}_{\mathrm{G+M}}(t) &  =\sum_{\substack{l\in\Gamma\cap\digamma
\\\Gamma,n\in\mathcal{R}}}\left\vert \mathcal{B},\Gamma,j_{l},i_{n}%
\right\rangle \left\langle \mathcal{B},\Gamma,j_{l},i_{n}\right\vert \hat
{U}_{\mathrm{G}}(t)\nonumber\\
&  \otimes\hat{U}_{\mathrm{M}\left\vert \mathrm{G}\right.  }^{\Gamma
(j_{l},i_{n})}(k_{l},F_{n},w_{n},t) \label{U-MG}%
\end{align}
As we noticed previously, the time-dependence of $\mathrm{S}_{\mathrm{G+M}}$ 
comes only from a phase factor that can be removed by
redefining $\left\vert \mathcal{B},\Gamma,j_{l},i_{n}\right\rangle $.

The picture underlying Eq.~\ref{U-MG}, similar to the above quantum-gate
interpretation of $\hat{U}_{\mathrm{G+M}}(t)$, is physically clear and
compelling: $\hat{U}_{\mathrm{G}}(t)$ creates from no spacetime a
superposition of the spin-network states and in the meanwhile, via programmed
entanglement operations $\hat{U}_{\mathrm{M}\left\vert \mathrm{G}\right.
}^{\Gamma(j_{l},i_{n})}(k_{l},F_{n},w_{n},t)$ generates matter states from no
matter, mamely:
\begin{align}
\hat{U}_{\mathrm{G}}(t)\left\vert \emptyset_{\mathrm{G}}\right\rangle  &
=\sum_{\substack{l\in\Gamma\cap\digamma\\\Gamma,n\in\mathcal{R}}%
}\mathrm{S}_{\mathrm{G+M}}[\Gamma,n,l;t]\left\vert \mathcal{B},\Gamma
,j_{l},i_{n}\right\rangle \nonumber\\
\hat{U}_{\mathrm{M}\left\vert \mathrm{G}\right.  }^{\Gamma(j_{l},i_{n}%
)}\left\vert \emptyset_{\mathrm{M}}\right\rangle  &  =\left\vert
(\psi,A),\Gamma,k_{l},F_{n},w_{n}\right\rangle \label{noMST}%
\end{align}
Obviously, $\left\vert (\psi,A),\Gamma,k_{l},F_{n},w_{n}\right\rangle $ also
defines a graph (the \textquotedblleft matter graph\textquotedblright) with
nodes and links. Then spacetime-matter entanglement is actually quantum
correlations between the spacetime graphs and the matter graphs. What is the
relation, if any, between the matter graphs and the Feynman graphs? This is
certainly an interesting future issue.

Needless to say, if the spin-network states indeed span the programming basis
as we assumed in this Section and the next Section, then physical picture
underlying our formulation is surely more transparent; many fruitful results
on, e.g., quantum geometry, are available. Regarding this, it remains to be
seen that the gravitational Hamiltonian in loop quantum gravity is indeed a
physical Hamiltonian. Meanwhile, the Euclidean entanglement action
$\mathcal{\hat{I}}_{\mathrm{Eu}}$ in this case can be given totally in terms
of the physical operators $\mathbf{\hat{A}}(\digamma)$ and $\mathbf{\hat{V}%
}(\mathcal{R})$ for geometry; later on we will show a particular example of
this situation (see Eq.~\ref{EuA-AV} below).

\section{Quantum information definition of dark energy}

The explicit form of $H_{\mathrm{M}\text{-}\mathrm{G}}(\hat{\psi},\hat{A};%
\mathcal{\hat{B}}_{a}^{i},\mathcal{\hat{E}}_{b}^{j})$ \cite%
{Thiemann,Rovelli-book,Thiemann-reg} shows that there are two kinds of the
gravity-matter coupling terms. Two terms are \textit{two-party couplings}
(denoted by $H_{\mathrm{2p}}^{\{\Gamma \}}$), each corresponding to the
interaction between gravity and matter fermions, or between gravity and
gauge fields; \textit{there is only one three-party coupling} term (denoted
by $H_{\mathrm{3p}}^{\{\Gamma \}}$) describing the interaction of gravity,
matter fermions, and gauge fields. As a result, in the Hilbert space of
matter only the three-party coupling term is responsible for entanglement
between matter fermions and gauge fields, as programmed by $\left\vert 
\mathcal{B},\Gamma ,j_{l},i_{n}\right\rangle $, while the two-party
couplings do not change the programmed matter entanglement. Meanwhile, in
the presence of matter, every link of the graph is labeled by the
irreducible $j$ representation of $SU(2)$ and the irreducible representation
of the gauge group, while fermions locate on the nodes. Intuitively, links
of the graph are the Faraday lines of forces \cite{Rovelli-book}; if there
is no link, there is no interaction. This intuitive picture motivates us to
rewrite $H_{\mathrm{M}\text{-}\mathrm{G}}^{\{\Gamma \}}=H_{\mathrm{2p}%
}^{\{\Gamma \}}+H_{\mathrm{3p}}^{\{\Gamma \}}$ as:
\begin{align}
H_{\mathrm{2p}}^{\{\Gamma \}}& =\sum_{\Gamma ,n\in \mathcal{R}}\left\vert
\Gamma ,j_{l}=\emptyset ,i_{n}\right\rangle \left\langle \Gamma
,j_{l}=\emptyset ,i_{n}\right\vert \nonumber\\
& \otimes H_{\mathrm{M}\left\vert \mathrm{G}%
\right. }^{\Gamma (i_{n})}(\hat{\psi},\hat{A})  \notag \\
H_{\mathrm{3p}}^{\{\Gamma \}}& =\sum_{\substack{ l\in \Gamma \cap \digamma 
\\ \Gamma ,n\in \mathcal{R}}}\left\vert \Gamma ,j_{l}\neq \emptyset
,i_{n}\right\rangle \left\langle \Gamma ,j_{l}\neq \emptyset
,i_{n}\right\vert \nonumber\\
& \otimes H_{\mathrm{M}\left\vert \mathrm{G}\right.
}^{\Gamma (j_{l}\neq \emptyset ,i_{n})}(\hat{\psi},\hat{A})  \label{p2p3}
\end{align}%
Here $H_{\mathrm{M}\left\vert \mathrm{G}\right. }^{\Gamma (i_{n})}(\hat{\psi}%
,\hat{A})=H_{\mathrm{Fermi}\left\vert \mathrm{G}\right. }^{\Gamma (i_{n})}(%
\hat{\psi})+H_{\mathrm{gauge}\left\vert \mathrm{G}\right. }^{\Gamma (i_{n})}(%
\hat{A})$; $H_{\mathrm{Fermi}\left\vert \mathrm{G}\right. }^{\Gamma (i_{n})}
$\ [$H_{\mathrm{gauge}\left\vert \mathrm{G}\right. }^{\Gamma
(i_{n})}$] is the Hamiltonian of matter fermions (gauge fields)
resulting from the two-party couplings $H_{\mathrm{2p}}^{\{\Gamma \}}$ in
the Hilbert space of matter.

As the programmed matter entangled state $\left\vert (\psi,A),\Gamma
,k_{l},\notag\right. \\ \left.F_{n},w_{n}\right\rangle $ is the energy eigenstate (see, e.g., Ref.~%
\cite{Rovelli-eigen} for simple examples of the energy eigenstates in loop
quantum gravity) of $H_{\mathrm{M}\left\vert \mathrm{G}\right.
}^{\Gamma(j_{l},i_{n})}(\hat{\psi},\hat{A})$, the corresponding eigenvalue
should have the following form 
\begin{equation}
E_{\mathrm{M}\left\vert \mathrm{G}\right. }^{\Gamma(j_{l},i_{n})}=\left\{ 
\begin{array}{c}
\mathbf{\bar{E}}_{\mathrm{M}\left\vert \mathrm{G}\right.
}^{\Gamma(i_{n})}(F_{n},w_{n})\text{ \ (without link exc.)} \\ 
E_{\mathrm{M}\left\vert \mathrm{G}\right.
}^{\Gamma(j_{l},i_{n})}(F_{n},w_{n};k_{l})\text{ \ (with link exc.)}%
\end{array}
\right.  \label{darkeg}
\end{equation}
where the \textquotedblleft node energy\textquotedblright\ $\mathbf{\bar{E}}%
_{\mathrm{M}\left\vert \mathrm{G}\right. }^{\Gamma(i_{n})}(F_{n},w_{n})$
[the \textquotedblleft link energy\textquotedblright\ $E_{\mathrm{M}%
\left\vert \mathrm{G}\right. }^{\Gamma(j_{l},i_{n})}(F_{n},w_{n};k_{l})$] is
the energy eigenvalue of $H_{\mathrm{M}\left\vert \mathrm{G}\right.
}^{\Gamma(i_{n})}(\hat{\psi},\hat{A})$ [$H_{\mathrm{M}\left\vert \mathrm{G}%
\right. }^{\Gamma(j_{l}\neq\emptyset,i_{n})}(\hat{\psi},\hat{A})$] related
to\ eigenstate $\left\vert
(\psi,A),\Gamma,k_{l}=\emptyset,\notag\right. \\ \left.F_{n},w_{n}\right\rangle $ without link
excitations [$\left\vert (\psi
,A),\Gamma,k_{l}\neq\emptyset,F_{n},w_{n}\right\rangle $ with link
excitations]. As noticed above, only those $\left\vert (\psi,A),\notag\right. \\ \left.\Gamma
,k_{l}\neq\emptyset,F_{n},w_{n}\right\rangle $ with link excitations have
programmed matter entanglement such that matter fermions and gauge fields
are mutually defined and measured. For those $\left\vert (\psi,A),\notag\right. \\ \left.\Gamma
,k_{l}=\emptyset,F_{n},w_{n}\right\rangle $ without link excitations, matter
fermions and gauge fields couple merely with gravity and as such, $\mathbf{%
\bar{E}}_{\mathrm{M}\left\vert \mathrm{G}\right.
}^{\Gamma(i_{n})}(F_{n},w_{n})$ must be \textit{dark energy}, which stems
from the two-party couplings and relates only to the volume excitations. In
other words, \textit{the ICQFT allows us to have a theoretical definition of
dark energy to be a kind of the bulk/volume energy}, while the surface/area
energy related to matter links is the \textquotedblleft visible
energy\textquotedblright.

Based on the above observations we can put $\left\vert \mathcal{B},(\psi
,A)\right\rangle $ into a superposition of a \textquotedblleft dark-energy
state\textquotedblright\ $\left\vert \mathrm{dark}\right\rangle $ and a
\textquotedblleft visible-energy state\textquotedblright\ $\left\vert 
\mathrm{visi}\right\rangle $, namely: 
\begin{equation}
\left\vert \mathcal{B},(\psi ,A)\right\rangle =p_{D}\left\vert 
\mathrm{dark}\right\rangle +q_{V}\left\vert \mathrm{visi}%
\right\rangle  \label{dv}
\end{equation}%
where $\left\vert p_{D} \right\vert ^{2}+\left\vert q_{V} \right\vert ^{2}=1$ and 
\begin{align}
\left\vert \mathrm{dark}\right\rangle & =\sum_{\Gamma ,n\in \mathcal{R}}%
\mathrm{S}_{\mathrm{G+M}}^{D}[\Gamma ,n]\left\vert \Gamma
,i_{n}\right\rangle \otimes \left\vert \Gamma ,F_{n},w_{n}\right\rangle  
\notag \\
\left\vert \mathrm{visi}\right\rangle & =\sum_{\substack{ l\in \Gamma \cap
\digamma  \\ \Gamma ,n\in \mathcal{R}}}\mathrm{S}_{\mathrm{G+M}}^{V}[\Gamma
,n,l]\left\vert \Gamma ,j_{l}\neq \emptyset ,i_{n}\right\rangle   \notag \\
& \otimes \left\vert \Gamma ,k_{l}\neq \emptyset ,F_{n},w_{n}\right\rangle
\label{dark-v}
\end{align}%
By definition, $\left\langle \mathrm{visi}\right. \left\vert \mathrm{dark}%
\right\rangle =0$. While for $\left\vert \mathrm{dark}\right\rangle $
spacetime and the fermion/gauge field are mutually defined and measured, the
fermion field and the gauge field are mutually defined and measured (i.e.,
visible to each other) for $\left\vert \mathrm{visi}\right\rangle $, as
programmed by gravity.

\section{Emergence of classical Einstein equation from spacetime-matter
entanglement}

Is the ICQFT given above a candidate theory of quantum gravity plus matter?
Here let us go further to illustrate one of the possible physical consequences
implied by a particular form of spacetime-matter entanglement, hoping to offer
a positive answer to this question. As the ICQFT is a concrete formalism of
interacting spacetime and matter, the state $\left\vert \mathcal{B}%
;A...,\psi...\right\rangle $ for spacetime and all matter contents contains
complete physical predictions of our Universe. Yet, how to calculate
explicitly the state, which generally includes infinite parameters, for physically interesting 
situations is a future challenge, unless we have the entanglement Hamiltonian or the Euclidean
entanglement action. Before doing any explicit calculations, one can consider
specific spacetime-matter entanglement that is well-based from other sides of
existing quantum gravity problems.

To this end, we specify dual spacetime-matter entanglement in
Eq.~\ref{Ent-GM} as:
\begin{equation}
\left\vert \mathrm{G},\mathrm{M}\right\rangle =\frac{1}{\sqrt{Z}}%
\sum\limits_{s}e^{-%
%TCIMACRO{\U{2124} }%
%BeginExpansion
\mathbb{Z}
%EndExpansion
_{s}/2}\left\vert \mathrm{G},s\right\rangle \otimes\text{$\left\vert
\mathrm{M},s\right\rangle $} \label{G-M}%
\end{equation}
where the Schmidt bases for the gravity and matter sectors are denoted
collectively by $\left\vert \mathrm{G},s\right\rangle $ and $\left\vert
\mathrm{M},s\right\rangle $, respectively. Here $Z$ is a normalization
constant and $%
%TCIMACRO{\U{2124} }%
%BeginExpansion
\mathbb{Z}
%EndExpansion
_{s}$ stands for the possible spectra for the gravitational and matter states.
Actually, gravity and matter are isospectral as their reduced density
operators read $\rho_{\mathrm{G}}=\frac{1}{Z}\sum\nolimits_{s}e^{-%
%TCIMACRO{\U{2124} }%
%BeginExpansion
\mathbb{Z}
%EndExpansion
_{s}}\left\vert \mathrm{G},s\right\rangle \left\langle \mathrm{G},s\right\vert
$ and $\rho_{\mathrm{M}}=\frac{1}{Z}\sum\nolimits_{s}e^{-%
%TCIMACRO{\U{2124} }%
%BeginExpansion
\mathbb{Z}
%EndExpansion
_{s}}\left\vert \mathrm{M},s\right\rangle \left\langle \mathrm{M},s\right\vert
$; for applications of pure-state entanglement like that in Eq.~\ref{G-M} in
a thermodynamic context, see \cite{Partovi}. Now let us suppose that
$\left\vert \mathrm{M},s\right\rangle $ ($\left\vert \mathrm{G},s\right\rangle
$) is an energy eigenvector of matter's Hamiltonian $H_{\mathrm{M}\left\vert
\mathrm{G}\right.  }$ (an area eigenvector of the area operator $\mathbf{\hat
{A}}$) with eigenvalue $E_{s}^{\mathrm{M}}$ ($\mathbf{A}_{s}^{\mathrm{G}}$)
such that: 
\begin{equation}
\beta E_{s}^{\mathrm{M}}=\tilde{\beta}\mathbf{A}_{s}^{\mathrm{G}}=
\mathbb{Z}_{s} \label{beta}%
\end{equation}
with $\beta$ and $\tilde{\beta}$\ being two constant factors.

If the matter field experiences a constant acceleration $a$, a Rindler horizon
appears due to the acceleration. As a uniformly accelerated observer in
Minkowski spacetime has no access to the states inside the Rindler horizon,
the reduced state for matter outside the Rindler horizon is a thermal state
characterized by the Unruh temperature $T_{U}=\frac{a\hslash}{2\pi c\kappa
_{B}}$, where $\kappa_{B}$\ and\ the speed of light $c$ are explicitly
included. This is known as the Unruh effect \cite{Unruh,QFT-cs,Unruh-RMP}
uncovered by a semi-classical analysis without quantizing gravity. Some recent
results \cite{FGP,Bianchi12,Bianchi13,Rovelli-Unruh,Perez,Smolin-Unruh}
studied the black-hole physics making use of the fact that the near-horizon
geometry of non-extremal black holes, as seen by a stationary observer, is
descriable by a local Rindler horizon. In these studies (e.g.,
\cite{Bianchi12,Bianchi13,Rovelli-Unruh,Perez,Smolin-Unruh}), entanglement
between the inside and the outside of the Rindler horizon is associated with
the black-hole entropy.

Instead of these previous results, here we consider whether or not the
spacetime-matter entangled state $\left\vert \mathrm{G},\mathrm{M}%
\right\rangle $\ could account for the Unruh effect. For this purpose, one can
identify $\beta=\frac{1}{\kappa_{B}T_{U}}$ in Eq.~\ref{beta} such that
$\rho_{\mathrm{M}}$\ is indeed a thermal state at the Unruh temperature
$T_{U}$. When there is a small perturbation to the whole system, the reduced
density operators will be $\rho_{\mathrm{G}}^{\prime}=\rho_{\mathrm{G}}%
+\delta\rho_{\mathrm{G}}$ and $\rho_{\mathrm{M}}^{\prime}=\rho_{\mathrm{M}%
}+\delta\rho_{\mathrm{M}}$. The change of the spacetime-matter entanglement
entropy at the first-order in $\delta\rho$ reads:
\begin{equation}
\delta\mathcal{E}_{\mathrm{GM}}=\tilde{\beta}\delta\left\langle \mathbf{\hat
{A}}\right\rangle \equiv\tilde{\beta}\delta\mathbf{A}=\beta\delta\left\langle
H_{\mathrm{M}\left\vert \mathrm{G}\right.  }\right\rangle \equiv\beta\delta
E_{\mathrm{M}\left\vert \mathrm{G}\right.  } \label{deltaS}%
\end{equation}
Here we have used Eq.~\ref{beta}, as well as the facts (see, e.g.,
\cite{Bianchi13,Rovelli-Unruh}) that $\delta\mathcal{E}_{\mathrm{GM}%
}=-\mathrm{tr}[\delta\rho_{\mathrm{G}}\ln\rho_{\mathrm{G}}]=-\mathrm{tr}%
[\delta\rho_{\mathrm{M}}\ln\rho_{\mathrm{M}}]$ and $\mathrm{tr}[\delta
\rho_{\mathrm{M}}]=0$. To be consistent with the Bekenstein-Hawking area law
\cite{Bekenstein,BekensteinPRD,Hawking}, one only needs to choose
$\tilde{\beta}$ to be a universal constant $\tilde{\beta}=\frac{1}{4\ell
_{P}^{2}}$, where the Planck length $\ell_{P}=\sqrt{G\hbar/c^{3}}$, such that:
\begin{equation}
\delta\mathcal{E}_{\mathrm{GM}}=\frac{\delta\mathbf{A}}{4\ell_{P}^{2}}%
\equiv\frac{\delta\mathbf{A}}{\mathbf{A}_{0}} \label{BH-area}%
\end{equation}
In particular, Eq.~\ref{deltaS} implies a relation:
\begin{equation}
\delta E_{\mathrm{M}\left\vert \mathrm{G}\right.  }=\frac{ac^{2}}{8\pi
G}\delta\mathbf{A} \label{FGP-EA}%
\end{equation}
which is identical in form to the Frodden-Gosh-Perez relation \cite{FGP}.

The celebrated work by Jacobson \cite{Jacobson} shows that the input of the
Unruh temperature and Eq.~\ref{deltaS} gives the classical Einstein
equation. This then means that our theory of quantum gravity has a correct
classical limit. A similar result was obtained within the context of loop
quantum gravity \cite{Rovelli-Unruh,Smolin-Unruh}. Moreover, if $\left\vert \mathrm{G},s\right\rangle
$\ is, instead, an energy eigenvector of gravity's Hamiltonian $H_{\mathrm{G}%
}$ (see Section {\bf 5}), following the above arguments and Eq.~\ref{HgHmg0} yields $\delta
\left\langle H_{\mathrm{G}}\right\rangle \equiv\delta E_{\mathrm{G}}%
=-\frac{ac^{2}}{8\pi G}\delta\mathbf{A}$ with the help of the energy-area
relation derived in loop quantum gravity
\cite{Bianchi12,Rovelli-Unruh,Smolin-Unruh}. In this case, while
one recovers the same Bekenstein-Hawking area law as in Eq.~\ref{BH-area},
the mean energy of gravity has to be identical to that of matter, but of
opposite sign such that \textit{the total mean energy of the whole
gravity+matter system is exactly zero}.

While the relations in Eqs.~\ref{BH-area} and \ref{FGP-EA} are formally
identical to those previous results
\cite{FGP,Bianchi12,Bianchi13,Rovelli-Unruh,Perez,Smolin-Unruh}, here the
physical picture is dramatically different because of different
entanglement involved. Moveover, note that in dual entanglement $\left\vert
\mathrm{G},\mathrm{M}\right\rangle $, the entangled states $\left\vert
\mathrm{M},s\right\rangle $ for the matter part depend of course on their
physical contents (i.e., matter species). However, as gravity universally
couples to matter via matter's energy-momentum tensor $T_{\mu}^{I}$, our
derivation of the relations in Eqs.~\ref{BH-area} and \ref{FGP-EA} does
not make use of any details on matter species. This explains the universal
independence of $\delta\mathcal{E}_{\mathrm{GM}}$ on matter species, known as
the species problem \cite{Bianchi12,Smolin-Unruh}.

\section{A universal spacetime-matter state of the Universe}

At first sight, it seems strange that Eq.~\ref{G-M} has no volume
excitations, unlike the Universe state $\left\vert \mathcal{B},(\psi
,A)\right\rangle $ in Eq.~(\ref{Ent-t}). This could be explained, following a
beautiful argument in Rovelli's book \cite{Rovelli-book}, by the fact that in
the presence of a horizon, only surface excitations are responsible for the
physics, especially for entropy counting. If this is indeed the case, the
volume excitations and the matter excitations programmed by them should be
absent or factorized away from the surface terms \textit{in some way}.

Here we would like to ask, besides Rovelli' argument, if there could be any
other fundamental reason explaining the absence (or presence) of the volume
excitations for entropy counting of the horizon. For this purpose, we can rewrite 
Eq.~\ref{G-M} as an area-matter entangled state: 
\begin{equation}
\left\vert \mathrm{G},\mathrm{M}\right\rangle _{\partial\Gamma}=\sum
_{\Gamma,l\in\Gamma\cap\digamma}\tfrac{e^{-\mathbf{A}(j_{l})/2\mathbf{A}_{0}}%
}{\sqrt{Z_{\partial\Gamma}}}\left\vert \mathrm{G},\partial\Gamma
,j_{l}\right\rangle \otimes\text{$\left\vert \mathrm{M},\partial\Gamma
,k_{l}\right\rangle $}\label{areaM}%
\end{equation}
Here the area states $\left\vert \mathrm{G},\partial\Gamma
,j_{l}\right\rangle $ are the eigenstates of $\mathbf{\hat{A}}(\digamma)$ with
eigenvalue $\mathbf{A}(j_{l})$, $\partial\Gamma$ means $\Gamma$'s boundary, i.e., its
intersections with the surface $\digamma$; $\left\vert \mathrm{M},\partial\Gamma
,k_{l}\right\rangle $ stands for the matter states programmed by $\left\vert
\mathrm{G},\partial\Gamma,j_{l}\right\rangle $. 
Inspecting the derivation of Eq.~\ref{deltaS}, one easily sees that the entanglement-area relation
\ref{BH-area} is universally valid for the area-matter entangled state in Eq.~\ref{areaM}.

Note that the relation $\delta\mathcal{E}_{\mathrm{GM}}=\frac{\delta
\mathbf{A}}{4\ell_{P}^{2}}$ can be regarded as a variational version of the
holographic principle \cite{tHooft,Susskind,holoRMP}, called the variational
holographic relation hereafter as it relates variations of two expectation
values ($\delta\mathcal{E}_{\mathrm{GM}}$ and $\delta\mathbf{A}$). Reversing
our reasoning that the specific form of $\left\vert \mathrm{G},\mathrm{M}%
\right\rangle _{\partial\Gamma}$ leads to the variational holographic
relation, we can take the variational holographic relation as a fundamental
principle that any theory of quantum gravity has to satisfy. Then it is
remarkable to see that the ICQFT, together with the variational holographic
relation, \textit{uniquely} determines the spacetime-matter entangled state as
given in (\ref{areaM}), in which the Schmidt coefficients $\mathrm{S}%
_{\mathrm{G+M}}$ can be specified. This encourages us to conjecture the
following spacetime-matter state:
\begin{align}
\left\vert \mathrm{Univ}\right\rangle  &  =\sum_{\substack{l\in\Gamma
\cap\digamma\\\Gamma,n\in\mathcal{R}}}\tfrac{e^{-\mathbf{V}(i_{n}%
)/2\mathbf{V}_{0}}\cdot e^{-\mathbf{A}(j_{l})/2\mathbf{A}_{0}}}{\sqrt
{Z_{\partial\Gamma}Z_{\Gamma}}}\left\vert \mathrm{G},\Gamma,i_{n}%
,j_{l}\right\rangle \nonumber\\
&  \otimes\text{$\left\vert \mathrm{M},\Gamma,F_{n},w_{n},k_{l}\right\rangle
$}\label{univ}%
\end{align}
where $Z_{\Gamma}$ is a new normalization constant and $\mathbf{V}_{0}$ a
volume constant to be determined. The Euclidean entanglement action
$\mathcal{\hat{I}}_{\mathrm{Eu}}^{(\Gamma+\partial\Gamma)}$ with respect to
$\left\vert \mathrm{Univ}\right\rangle $ is
\begin{equation}
\mathcal{\hat{I}}_{\mathrm{Eu}}^{(\Gamma+\partial\Gamma)}=\frac{\mathbf{\hat
{A}}(\digamma)}{\mathbf{A}_{0}}+\frac{\mathbf{\hat{V}}(\mathcal{R}%
)}{\mathbf{V}_{0}} \label{EuA-AV}%
\end{equation}
which is purely geometric.

Now let us explain how we can arrive at $\left\vert \mathrm{Univ}\right\rangle
$. To this end, we return to the quantum-gate interpretation of $\hat
{U}_{\mathrm{G+M}}(t)$. By this interpretation, the dynamical evolution
resulting in $\left\vert \mathrm{Univ}\right\rangle $ [$\left\vert
\mathcal{B},(\psi,A)\right\rangle $] is exactly the computing process of an
information-complete quantum computer defined in Ref.~\cite{ICQT} if we use
$\Gamma_{T}$ ($T=0,1,2,...$) to label the computing steps, which actually
defines discrete time. Note that $\left\vert \mathcal{B},\Gamma,j_{l}%
,i_{n}\right\rangle $ [$\left\vert (\psi,A),\Gamma,k_{l},F_{n},w_{n}%
\right\rangle $] in Eq.~\ref{Ent-t} is the energy eigenstate of
$H_{\mathrm{G}}$ [$H_{\mathrm{M}\left\vert \mathrm{G}\right.  }^{\Gamma
(j_{l},i_{n})}$] according to the preceding Section. $\Gamma$ thus labels the
total energy of spacetime or matter for a given graph. The
information-complete quantum computing proceeds from $T=0$ (the empty state
$\left\vert \emptyset\right\rangle $) and consumes matter of increasing
energies step by step, resulting in expanded spacetime and more matter
described by $\left\vert \mathrm{Univ}\right\rangle $. During expanding
spacetime and creating matter, the spacetime and matter graphs grow up and get
more entangled. In this process spacetime and matter \textquotedblleft
borrow\textquotedblright\ energies from each other while keeping the total
energy of the trinary fields exactly zero.

If spacetime-matter entanglement has a universal form shown in Eq.~\ref{univ}, 
the total entanglement entropy is the sum of entanglement entropies for
nodes and for links---the additivity of volume and area entanglement
entropies; in particular, the variational holographic relation is modified as:
\begin{equation}
\delta\mathcal{E}_{\mathrm{GM}}^{(\Gamma+\partial\Gamma)}=\frac{\delta
\mathbf{A}}{\mathbf{A}_{0}}+\frac{\delta\mathbf{V}}{\mathbf{V}_{0}}
\label{eav}%
\end{equation}
which is also of a universal form. In other words, the information-complete
quantum computing for spacetime and matter in $\left\vert \mathrm{Univ}%
\right\rangle $ results in a monotone increasing, by a fixed and universal
amount $\delta\mathcal{E}_{\mathrm{GM}}^{(\Gamma+\partial\Gamma)}$ for each
computing step ($\Gamma_{T}\rightarrow\Gamma_{T+1}$ for large enough $T$), of
the spacetime-matter entanglement entropy. \textit{This monotonically
increasing entanglement entropy thus defines an arrow of time}. If we understand
$\left\vert \mathrm{Univ}\right\rangle$ in a cosmological context, such an arrow of time
is something like the usual cosmological arrow of time. As we use the discrete computing step 
to label time, it is hard to relate such a discrete time with the time in the dynamical equation.
Equation~\ref{eav} generalizes the variational holographic relation given in
Eq.~\ref{BH-area}, which is approximately valid for large $\mathbf{V}_{0}$.

Obviously, $\left\vert \mathrm{G},\mathrm{M}\right\rangle _{\partial\Gamma}$
describes a Universe where matter can only entangle with quantized surface. In
other words, the area-matter entangled state encodes complete information of
physical predictions for the strictly holographic Universe, where the
programming basis has to be switched from \{$\left\vert \Gamma,j_{l}%
,i_{n}\right\rangle $\} to \{$\left\vert \mathcal{B},\partial\Gamma
,j_{l}\right\rangle $\}. Such a truncation of the spin-network Hilbert space
can be done, e.g., by taking the node (volume) degrees of freedom of
spin-networks as pure gauge \cite{cograin}. However, as we already noted, the
node energy contributes to the dynamics of the whole trinary fields.
Consequently, \textit{our Universe is not strictly holographic due to the
presence of dark energy}. Similarly to Eq.~\ref{beta}, we have $\beta
_{L}E_{\mathrm{M}\left\vert \mathrm{G}\right.  }^{\Gamma(j_{l},i_{n})}%
+\beta_{N}\mathbf{\bar{E}}_{\mathrm{M}\left\vert \mathrm{G}\right.  }%
^{\Gamma(i_{n})}=\mathbf{A}(j_{l})/\mathbf{A}_{0}+\mathbf{V}(i_{n}%
)/\mathbf{V}_{0}$, where $\beta_{L}$ and $\beta_{N}$ are two constant factors
related to the link energy and the node energy, respectively. Consequently:
\begin{equation}
\beta_{L}\delta E_{\mathrm{M}\left\vert \mathrm{G}\right.  }^{\Gamma}%
+\beta_{N}\delta\mathbf{\bar{E}}_{\mathrm{M}\left\vert \mathrm{G}\right.
}^{\Gamma}=\frac{\delta\mathbf{A}}{\mathbf{A}_{0}}+\frac{\delta\mathbf{V}%
}{\mathbf{V}_{0}} \label{eeva}%
\end{equation}
This relation generalizes the variational energy-area relation in
Eq.~\ref{FGP-EA}.

\section{The cosmological constant term}

Now let us consider the application of the results, given in the above
Section, to the problem of the cosmological constant \cite{cc-Weinberg}. Note
that in loop quantum gravity, the Hamiltonian related to the cosmological
constant term reads (after restoring $G$, $\hbar$, and $c$):
\begin{equation}
H_{\Lambda}=-\frac{\hbar c\Lambda}{8\pi\ell_{P}^{2}}\int_{\mathcal{R}}%
d^{3}x\sqrt{\det g} \label{Hcosmoc}%
\end{equation}
where the 3-metric $g_{ab}=e_{a}^{i}e_{b}^{i}$ and $\int_{\mathcal{R}}%
d^{3}x\sqrt{\det g}$ is classically the total volume of the region
\cite{Thiemann,Thiemann-reg}. If we assume that $H_{\Lambda}$ after
quantization is contributed solely by the dark energy defined by our theory,
we would have $\beta_{N}\delta\mathbf{\bar{E}}_{\mathrm{M}\left\vert
\mathrm{G}\right.  }^{\Gamma}=\beta_{N}\frac{\hbar c\Lambda}{8\pi\ell_{P}%
^{2}}\delta\mathbf{V=}\delta\mathbf{V}/\mathbf{V}_{0}$ (Note that the mean
value of $H_{\Lambda}$ should be identical to the dark energy, but of opposite
sign). This allows us to determine $\mathbf{V}_{0}$ as $\mathbf{V}_{0}%
^{-1}=\beta_{N}\frac{\hbar c\Lambda}{8\pi\ell_{P}^{2}}$, keeping $\beta_{N}$
undetermined. If $\left\vert \mathrm{Univ}\right\rangle $ describes an
expanding Universe, which expands at a constant acceleration $a_{E}$, a
natural conjecture might be $\beta_{N}^{-1}=\frac{\hbar a_{E}}{2\pi c}%
\equiv\frac{\hbar}{2\pi t_{E}}\equiv\kappa_{B}T_{E}$ such that:
\begin{equation}
\mathbf{V}_{0}=\frac{4\ell_{P}^{2}}{ct_{E}\Lambda} \label{vbeta}%
\end{equation}
$\mathbf{V}_{0}$ can indeed be very large if $\frac{1}{ct_{E}\Lambda}$ is a
length scale comparable to the Hubble length $\sim\sqrt{1/\Lambda}$. Thus, for
the expanding Universe described by $\left\vert \mathrm{Univ}\right\rangle $,
the cosmological constant term in Einstein's equation could be attributed to
the dark/node energy, related to the volume quanta, defined by our theory.

Recall the thermodynamic relation $dE=TdS-Pd\mathbf{V}$ relating energy $E$,
entropy $S$, temperature $T$, pressure $P$, and volume $\mathbf{V}$. For this
relation to hold in the present case, we need to require $\beta_{L}=\beta_{N}%
$, which allows us to define the total \textquotedblleft entanglement
energy\textquotedblright\ $\Xi_{\Gamma}\equiv E_{\mathrm{M}\left\vert
\mathrm{G}\right.  }^{\Gamma}+\mathbf{\bar{E}}_{\mathrm{M}\left\vert
\mathrm{G}\right.  }^{\Gamma}$ (see the definition of the entanglement
Hamiltonian in Eq.~\ref{eH}). From Eq.~\ref{eeva} we then obtain quite
similarly a thermodynamic relation:
\begin{equation}
\delta\Xi_{\Gamma}=T_{E}\delta S-P_{U}\delta\mathbf{V} \label{ee-STV}%
\end{equation}
provided that:
\begin{align}
\delta S  &  =\frac{\kappa_{B}\delta\mathbf{A}}{\mathbf{A}_{0}},\nonumber\\
P_{U}  &  =-\frac{\kappa_{B}T_{E}}{\mathbf{V}_{0}}=-\frac{\hbar c\Lambda
}{8\pi\ell_{P}^{2}}=-\frac{c^{4}\Lambda}{8\pi G} \label{SPV}%
\end{align}
It is ready to see that the extra term $\delta\mathbf{V/V}_{0}$ in
Eq.~\ref{eeva} gives a negative pressure $P_{U}$, which is a universal
constant given by three fundamental constants $c$, $\Lambda$, and $G$. The
universal negative pressure is believed to expand, at the constant
acceleration $a_{E}$, our Universe. This picture is consistent with current
understanding of the standard cosmology and here stems directly from our
theory structure. The physical significance of Eq.~\ref{SPV} is transparent:
\textit{The total entanglement energy }$\Xi_{\Gamma}$\textit{ is the physical
energy}, which consists of two parts---the visible/holographic energy related
to the variational holographic relation $T_{E}\delta S=\frac{\kappa_{B}%
T_{E}\delta\mathbf{A}}{\mathbf{A}_{0}}$\ (see Eq.~\ref{FGP-EA}) and the dark
energy related to $\left\vert P_{U}\right\vert \delta\mathbf{V}$. Note that
$\sqrt[3]{\mathbf{V}_{0}}$ define a crossover length scale in between
$\ell_{P}$ and $\sqrt{1/\Lambda}$ for the holographic and dark energies.

In Jacobson's thermodynamic argument to derive the classical Einstein equation
\cite{Jacobson}, the cosmological constant still remains to be a free
parameter. If we make use of Eq.~\ref{eav}, rather than Eq.~\ref{deltaS},
in such an argument, our above discussion on the universal relation between
entanglement entropy and geometry (area and volume) has actually fixed the
cosmological constant term in Einstein's equation. Therefore, we readily see
that the spacetime-matter entangled state $\left\vert \mathrm{Univ}%
\right\rangle $ in Eq.~\ref{univ} provides more complete information of our
Universe than $\left\vert \mathrm{G},\mathrm{M}\right\rangle _{\partial\Gamma
}$ in Eq.~\ref{areaM}.

The conceptual application of the above results is profound. What we have done
in this and the above two Sections is to consider two particular forms of
spacetime-matter entangled states ($\left\vert \mathrm{G},\mathrm{M}%
\right\rangle _{\partial\Gamma}$ and $\left\vert \mathrm{Univ}\right\rangle $)
and their physical consequences. While $\left\vert \mathrm{G},\mathrm{M}%
\right\rangle _{\partial\Gamma}$\ is consistent with the variational
holographic relation, $\left\vert \mathrm{Univ}\right\rangle $ results in a
more realistic relation between entanglement entropy and geometry. It is
$\left\vert \mathrm{Univ}\right\rangle $ that allows us to determine the
cosmological constant term in Einstein's equation. Reversing the reasoning
supports $\left\vert \mathrm{Univ}\right\rangle $ as a reliable quantum state
of the Universe and moreover, the fact that our Universe is not strictly holographic.

\section{Quantum state of a Schwarzschild black hole}

Let us present one more application of the dual entanglement structure of our
theory. We consider the limit on the information content of a spacetime region
$\mathcal{R}$ associated with a surface $\partial\mathcal{R}$ of area
$\mathbf{A}$. Let us identify the spacetime region (elementary fermions and
gauge fields associated within the spacetime region) as the $\mathcal{P}$
($\mathcal{SA}$) system. Then the $\mathcal{P}$-$\mathcal{SA}$ measurability
and the programmed measurability $\mathcal{SA}\left\vert _{\mathcal{P}%
}\right.  $ defined in the ICQT \cite{ICQT} demand that $D_{\mathcal{A}%
}=D_{\mathcal{S}}=D$ and \textit{maximally} $D_{\mathcal{P}}=D^{2}$, i.e., the
dimensions (denoted by $D_{\mathcal{A},\mathcal{S},\mathcal{P}}$) of the three
systems are all limited and related. These facts immediately lead to an
obvious relation:
\begin{equation}
\mathcal{E}_{\mathcal{P}(\mathcal{SA})}\leq\ln D_{\mathcal{P}} \label{strong}
\end{equation}
namely, spacetime-matter ($\mathcal{P}$-$\mathcal{SA}$) entanglement, as
quantified by $\mathcal{E}_{\mathcal{P}(\mathcal{SA})}$ (the entropy of
$\mathcal{P}$ or $\mathcal{SA}$), is limited in our picture. Here, the
equality applies only to the case of maximal $\mathcal{P}$-$\mathcal{SA}$ entanglement.

Remarkably, loop quantum gravity can give a complete spectrum of the area operator, see
Refs.~\cite{Thiemann,Rovelli-book,Chiou}). Using the area spectrum and microstate counting \cite{Rovelli-book}
applied to a Schwarzschild black hole of surface area
$\mathbf{A}$, Eq.~\ref{strong} becomes
\begin{equation}
\mathcal{E}_{\mathcal{P}(\mathcal{SA})}\leq\ln D_{\mathcal{P}}=\frac{\mathbf{A}}{4\ell_{P}^{2}}
\label{holo}
\end{equation}
This is exactly what the holographic principle \cite{tHooft,Susskind,holoRMP} implying that the
information content is limited merely by the surface. In the context of the ICQFT, the holographic
principle arises as a direct consequence of area-matter entanglement. From such a
strong and universal limit on the allowed states of the trinary system as
imposed by the ICQFT, it is ready to see, once again, that the restriction on the description of the trinary
fields imposed by the information-completeness is much stronger than our
current field-theoretical description.

Now let us apply the above argument to the Schwarzschild black hole of surface area
$\mathbf{A}$. As is now widely accepted, the black hole saturates
\cite{Susskind-BH} the entanglement bound in Eq.~\ref{holo}. This single fact is so special that it is enough
for us to infer the global state of the black hole, namely, the black hole must be a \textit{maximally information-complete} quantum
system with the maximal area-matter entanglement:
\begin{equation}
\left\vert \mathrm{BH},\mathcal{P(SA)}\right\rangle _{\partial\Gamma}%
=\tfrac{1}{\sqrt{D_{\mathcal{P}}^{(L)}}}\sum_{\Gamma,l\in\Gamma\cap
\partial\mathcal{R}}\left\vert \mathcal{P},\partial\Gamma,j_{l}\right\rangle
\otimes\text{$\left\vert \mathcal{SA},\partial\Gamma,k_{l}\right\rangle $%
} \label{BHstate}%
\end{equation}
when only the surface of the black hole is concerned. Here $D_{\mathcal{P}%
}^{(L)}=2^{\mathbf{A}/{4\ell_{P}^{2}}}$ (see, e.g., Ref.~\cite{Rovelli-book}) is 
the dimensions of the link states $\left\vert \mathcal{P}%
,\partial\Gamma,j_{l}\right\rangle $; all $D_{\mathcal{P}}^{(L)}$ matter link
states $\left\vert \mathcal{SA},\partial\Gamma,k_{l}\right\rangle $ are also
maximally entangled and span an orthonormal basis in the matter sector. In
this case the entanglement entropy of the black hole is $\mathcal{E}%
_{\mathcal{P}(\mathcal{SA})}^{\mathrm{BH,}\partial\Gamma}=-\mathrm{tr}%
[\rho_{\mathcal{P}}^{\{l\}}\ln\rho_{\mathcal{P}}^{\{l\}}]=\ln D_{\mathcal{P}%
}^{(L)}=\frac{\mathbf{A}}{4\ell_{P}^{2}}$, where $\rho_{\mathcal{P}}%
^{\{l\}}=\mathrm{tr}_{\mathcal{SA}}(\left\vert \mathrm{BH},\mathcal{P(SA)}%
\right\rangle _{\partial\Gamma}\left\langle \mathrm{BH},\mathcal{P(SA)}%
\right\vert )=\frac{1}{D_{\mathcal{P}}^{(L)}}\sum_{\Gamma,l\in\Gamma
\cap\partial\mathcal{R}}\left\vert \mathcal{P},\partial\Gamma,j_{l}%
\right\rangle \left\langle \mathcal{P},\partial\Gamma,j_{l}\right\vert $; see,
as a comparison, Eq.~\ref{areaM}, in which the area-matter state is not
maximally entangled and leads to the entanglement entropy less than $\ln
D_{\mathcal{P}}^{(L)}$.

However, as we emphasized above, \textit{the variational holographic relation
is violated by our Universe} and has to be modified into the form given by
Eq.~\ref{eav}. Thus, we must consider $\partial\mathcal{R}$ and
$\mathcal{R}$ to give the global state of the black hole. As such,
$D_{\mathcal{P}}$ in Eq.~\ref{strong} is the total dimensions of the
spacetime state related to $\partial\mathcal{R}$ and $\mathcal{R}$, and the
black hole should be described by the total quantum state that is maximally
information-complete in both volume and surface degrees of freedom, namely:
\begin{align}
\left\vert \mathrm{BH},\mathcal{P(SA)}\right\rangle _{\Gamma} &  =\tfrac
{1}{\sqrt{D_{\mathcal{P}}^{(L)}D_{\mathcal{P}}^{(N)}}}\sum_{\substack{l\in
\Gamma\cap\partial\mathcal{R}\\\Gamma,n\in\mathcal{R}}}\left\vert
\mathcal{P},\Gamma,i_{n},j_{l}\right\rangle \nonumber\\
&  \otimes\text{$\left\vert \mathcal{SA},\Gamma,F_{n},w_{n},k_{l}\right\rangle
$}\label{bh-bulk}%
\end{align}
where $D_{\mathcal{P}}^{(N)}$ is the dimensions of the node states. Further
consideration of entanglement entropy with respect to 
$\left\vert \mathrm{BH},\mathcal{P(SA)}\right\rangle _{\Gamma}$
will be given elsewhere.

Maximal entanglement has an intriguing property called \textquotedblleft
monogamy\textquotedblright\ \cite{mono}: If two parties are maximally
entangled, then they cannot be entangled with any third party. Let us discuss
a possible application of this \textquotedblleft
non-shareability\textquotedblright\ of maximal entanglement in the present
context. As we inferred, the black hole is maximally entangled in dual form.
Then the monogamy of maximal entanglement implies that there is no way of
extracting any information, via interactions, from the black hole. Namely, the
black hole is \textquotedblleft information-black\textquotedblright. As such,
dynamical evolution of the black hole will be in some sense \textquotedblleft
frozen\textquotedblright\ from the trinary fields, namely, it is an
\textquotedblleft\textit{entanglement death}\textquotedblright\ of matter and
spacetime. However, the presence of the black hole in spacetime is detectable
as it \textit{defines} spacetime and can also absorb matter to grow up its
entanglement. Such a picture on black holes seems to be in accordance with our
intuition on what \textit{is} a black hole, especially in the framework of
ICQFT. However, it is quite different from our current understanding
\cite{Bekenstein,BekensteinPRD,Hawking,Thiemann} based on classical general
relativity, thermodynamic argument, and QFT in curved
classical spacetime.

Note that in $\left\vert \mathrm{BH},\mathcal{P(SA)}\right\rangle _{\Gamma}$,
there are both the volume and surface quanta (as well as the programmed
matter) inside the Schwarzschild black hole. In this case, the interior of the
black hole must be factorized away from quantum states for degrees of freedom
outside the black hole as $\left\vert \mathrm{BH},\mathcal{P(SA)}\right\rangle
_{\Gamma}$ is maximally dual-entangled---external matter fields can entangle
only with surface excitations near but outside the horizon, in a state
approximately given by $\left\vert \mathrm{G},\mathrm{M}\right\rangle
_{\partial\Gamma}$ as typically $\mathbf{V}_{0}$ is much larger than the
quantized volume. This gives a physical explanation validating the argument
on the derivation of the variational holographic relation without the volume
contribution in the presence of a horizon. Moreover, as $\left\vert
\mathrm{BH},\mathcal{P(SA)}\right\rangle _{\Gamma}$ is maximally
information-complete, but regular, for the Schwarzschild black hole
\textit{the singularity problem disappears for the inferred state }$\left\vert
\mathrm{BH},\mathcal{P(SA)}\right\rangle _{\Gamma}$. Such a maximal
entanglement possessed by the Schwarzschild black hole might be an ideal
\textquotedblleft resource\textquotedblright\ for quantum information
processing with matter and spacetime.

What about the black-hole information paradox in the ICQFT? The successful
account of the Bekenstein-Hawking entropy (as well as the cosmological
constant term) totally in terms of spacetime-matter entanglement convinces us
the elimination of this paradox within the ICQFT. Here, matter's states seem
to be thermal not because some modes of matter fields are thrown into a black
hole which has \textquotedblleft no hair\textquotedblright\ and thus destroys
information about collapsing matter. Rather, all information of the whole
system is coherently kept in dual entanglement and the thermality of matter's
states stems from an information-incomplete description, i.e., artificially
tracing out the spacetime degrees of freedom within dual
entanglement---\textit{The black hole as an information-complete system is not
thermal; it thus does not evaporate and never destroys information}.
\textit{Nowhere is a non-unitary evolution allowed in the ICQFT}. Recall that Hawking's radiation
was derived from quantum field theory in classical curved spacetime. By contrast, in the 
unitary and information-complete description of nature given in this work, there is 
simply no Hawking radiation and information is definitely conservative so that there is no room for 
any information loss. In essence, the black-hole information paradox originates from the
information-incompleteness of current quantum description. As a comparison,
loop quantum gravity helps to solve the singularity problem, but the information-loss problem of
black holes becomes worse \cite{Bojowald}.

\section{Summary and outlook}

To summarize, we have introduced, very briefly as a start, the
ICQFT, as quantum entanglement dynamics of
spacetime and matter, which describes elementary matter fermions, their gauge
fields, and gravity as an indivisible trinity, hoping to provide a coherent picture of
unifying spacetime and matter. The fact that this is indeed possible could be
regarded as a support on our previous argument on the information-complete
quantum theory. Complete information of the trinary fields is encoded in the
dual entanglement---spacetime-matter entanglement and matter-matter (matter
fermions and their gauge fields) entanglement. Thus, in terms of entanglement,
both spacetime and matter are unified as information. Here,
entanglement is universal just like that gravity is universal; the universal
entanglement glues spacetime and matter and is thus the building block of the
world. Working in dual entanglement formalism of the trinary fields has an
obvious advantage in that all constraint conditions are automatically solved
and all predictions of the theory are exactly physical (neither more nor less)
and explicit. We give a consistent framework of the dual dynamical evolution of the
trinary fields, which accounts for the two pieces of Einstein's equation.

Any reliable theory of quantum gravity must, first of all, make progress on
existing conceptual problems, which remain to be \textquotedblleft a major obstacle
for the final construction of a quantum theory of gravity and its application
to cosmology.\textquotedblright\ \cite{Kiefer} As a concrete formulation of
quantum theory for spacetime and matter, several conceptual progresses, we
believe, have been made within our theory as follows.

\begin{description}
\item[No probability and no observer] Our information-complete quantum
description does not rely in any way on the concepts of probability and
observer, or any related classical concepts in its own formulation---A
probability description appears only as an information-incomplete and
approximate description of nature; \textit{our Universe is self-defining and
self-explaining via its trinary constituents}, but not defined and explained
via any external observers, as shown previously in a new quantum structure
beyond current quantum theory \cite{ICQT}. This eliminates the conceptual
obstacle of applying conventional quantum theory to cosmology. The conjectured
quantum state of the Universe leads to encouraging results.

\item[Dark energy] The ICQFT provides a quantum information definition of the
mysterious dark energy, which stems from the two-party couplings within the
spacetime-matter Hamiltonian $H_{\mathrm{M}\text{-}\mathrm{G}}$. For such
couplings, matter fermions and their gauge fields interact merely with gravity
and the related energy is thus dark energy, shown to be a kind of the
node/volume energy.

\item[Problem of time and time's arrow] While the overall quantum state of
spacetime and matter is timeless, its particular entanglement structure allows
us to define separately time evolutions for spacetime and for matter. Related
to this and the basic property of entanglement, our theory implies an
entanglement-induced arrow of time. The Universe, as an information-complete
quantum computer, has a monotonically increasing entanglement entropy,
defining also an arrow of time.

\item[The variational holographic relation] It is actually violated by our
Universe\textit{. }Based on a particular form of area-matter entanglement, we
derived the variational holographic relation (i.e., the variational form of
the entropy-area law), which can lead to the classical Einstein equation
(Jacobson's thermodynamic argument), but leaving the cosmological constant as
a free parameter. A more general form of spacetime-matter entanglement results
in the universal relation between entanglement entropy and geometry (see
Eq.~\ref{eav}), which modifies the variational holographic relation].
Therein, the extra term ($\delta\mathbf{V/V}_{0}$) was argued to be
responsible for the cosmological constant term in Einstein's equation. This
latter fact also confirms the above-mentioned picture of dark energy as a kind
of the volume energy. In a thermodynamic argument, dark energy corresponds to
$\left\vert P_{U}\right\vert \mathbf{V}$, where $P_{U}=-\frac{c^{4}\Lambda
}{8\pi G}$ is a universal negative pressure expanding our Universe with the
volume $\mathbf{V}$.

\item[Quantum black hole] For a Schwarzschild black hole, we infer that it is
the maximally information-complete quantum system with maximal dual
entanglement, whose monogamy enables a conceptually clear understanding of the
black hole. As there is no room for non-unitary evolution, there is no
information-loss paradox in our information-complete trinary description of nature.
\end{description}

We would like to emphasize that, as usual quantum mechanics, current QFT 
is also information-incomplete and describes elementary fermion
fields or/and gauge fields as isolated, physical entities. This description
leads to interpretational difficulties such as the black-hole information
paradox and physical meaning of field quanta in curved spacetime. In the
ICQFT, however, a dramatically different picture arises. Here spacetime
(gravity) and matter are mutually defined and entangled---no spacetime implies
no matter, and vice versa. As programmed by spacetime, elementary fermion
fields and their gauge fields are likewise mutually defined and entangled;
either of them alone cannot be information-complete. In some sense, it is the
quantum version of Einstein's gravity that completes the picture. The ICQFT,
free of those interpretational difficulties or paradox that we encountered in
conventional QFT, calls for a radical change of our current
understanding on spacetime, matter, information and reality, as well as their
relations. In the ICQFT, which deals with a self-explaining Universe,
spacetime and matter are unified into information (entanglement) of direct
physical reality. Here, \textit{it is not the constituent parts (elementary
matter fermions and gauge fields, as well as spacetime) of the Universe, but
rather their relations (i.e., entanglement) that are physical.}

We have shown thus far that for both quantum mechanical systems and quantum
fields, information-complete description of trinary systems shares common
features such as dual entanglement, dual dynamics, and exclusion of any
classical concepts like probability description. The mere possibility of
achieving this is itself a surprise and conceptually appealing. Compared to
current quantum theory and general relativity, dual dynamics pertaining to
dual entanglement of the trinary description is a new feature. Rather than
Wheeler's famous coinage on general relativity---\textquotedblleft Space tells
matter how to move and matter tells space how to curve\textquotedblright%
\ \cite{MTZ}, here we would like to say that \textit{spacetime-matter
entanglement moves matter and curves spacetime quantum mechanically; even
more, it defines spacetime and matter}. These claims stem from the fact that
spacetime-matter entanglement leads to a correct classical limit, namely,
Einstein's equation.

Like existing approaches to quantum gravity, there are too many open questions
in the framework of the ICQFT, including more physical consequences implied by
the information complete trinary description, the relation between the ICQFT
and conventional QFT, and so on. If this work serves as a
start to stimulate someone to take into account seriously and to work out more
consequences of our information-complete trinary description of nature, it is
exactly the author's hope.

\bigskip
\noindent\textbf{Declaration of competing interest}\\
The author declares that he has no conflicts of interest in this work.\\

\bigskip
\noindent\textbf{Acknowledgments}\\
I am grateful to Chang-Pu Sun for bringing Ref. \cite{Tipler} into my attention. I also acknowledge University of Science and Technology of China, where the work was initiated. This research was funded by the Fundamental Research Funds for the Central Universities under grant number 020414380182.\\

%%%%%%%%%%%%%%%%%%%%%%%%%%%%%%%%%%%%%%%%%%%%%%%%%%%%%%%
%%% Supplements. ????????, ????
%%%%%%%%%%%%%%%%%%%%%%%%%%%%%%%%%%%%%%%%%%%%%%%%%%%%%%%
%\Supplements{}

%%%%%%%%%%%%%%%%%%%%%%%%%%%%%%%%%%%%%%%%%%%%%%%%%%%%%%%
%%% Reference section. ?ο?????
%%% citation in the content using "some words~\cite{1,2}".
%%% ~ is needed to make the reference number is on the same line with the word before it.
%%%%%%%%%%%%%%%%%%%%%%%%%%%%%%%%%%%%%%%%%%%%%%%%%%%%%%%


\begin{thebibliography}{999}
% Reference 1
\bibitem {Thiemann}T. Thiemann, Lectures on loop quantum gravity,
Lecture Notes in Physics \textbf{631}  (2003) 41-135.

\bibitem {AC-nature}G. Amelino-Camelia, Quantum theory's last
challenge, Nature \textbf{408} (2000) 661-664.

\bibitem {quanA1}C. Rovelli, L. Smolin, Discreteness of area and
volume in quantum gravity, Nucl. Phys. B \textbf{442} (1995) 593-619.

\bibitem {quanA2}C. Rovelli, Black hole entropy from loop quantum
gravity, Phys. Rev. Lett. \textbf{77} (1996) 3288-3291.

\bibitem {quanA3}A. Ashtekar, J. Baez, A. Corichi, A. Krasnov,
Quantum geometry and black hole entropy, Phys. Rev. Lett.
\textbf{80} (1998) 904-907.

\bibitem {loop25}C. Rovelli, Loop quantum gravity: the first 25
years, Class. Quantum Grav. \textbf{28} (2011) 153002.

\bibitem {Rovelli-book}C. Rovelli, Quantum Gravity (Cambridge Univ.
Press, Cambridge, 2004).

\bibitem {Thiemann-reg}T. Thiemann, Quantum spin dynamics (QSD): V.
Quantum gravity as the natural regulator of the Hamiltonian constraint of
matter quantum field theories, Class. Quantum Grav. \textbf{15} (1998) 1281-1314.

\bibitem {Chiou}D.-W. Chiou, Loop quantum gravity, Int. J. Mod. Phys.
D \textbf{24} (2015) 1530005.

\bibitem {EPR}A. Einstein, B. Podolsky, N. Rosen, Can
quantum-mechanical description of physical reality be considered complete?
Phys. Rev. \textbf{47} (1935) 777-780.

\bibitem {Bohr}N. Bohr, Can quantum-mechanical description of physical
reality be considered complete? Phys. Rev. \textbf{48} (1935) 696-702.

\bibitem {PBR}M.F. Pusey, J. Barrett, T. Rudolph, On the reality
of the quantum state, Nature Phys. \textbf{8} (2012) 475-478.

\bibitem {vonN-book}J. von Neumann, Mathematical Foundations of
Quantum Mechanics, Vol. 2 (Princeton Univ. Press, Princeton, New Jersey, 1996).

\bibitem {WZ-book}J.A. Wheeler, W.H. Zurek, eds., Quantum Theory
and Measurement (Princeton Univ. Press, Princeton, New Jersey, 1983).

\bibitem {Zurek}W.H. Zurek, Decoherence, einselection, and the quantum
origins of the classical, Rev. Mod. Phys. \textbf{75} (2003) 715-775.

\bibitem {Sun}C.-P. Sun, X.-X. Yi, X.-J. Liu, Quantum dynamical
approach of wavefunction collapse in measurement progress and its application
to quantum Zeno effect, Fortschr. Phys. \textbf{43} (1995) 585-612.

\bibitem {Susskind-BH}L. Susskind, The paradox of quantum black
holes, Nature Phys. \textbf{2}, 665-677 (2006).

\bibitem {Unruh}W.G. Unruh, Notes on black hole evaporation, Phys.
Rev. D \textbf{14} (1976) 870-892.

\bibitem {QFT-cs}N.D. Birrell, P.C.W. Davies, Quantum Fields in
Curved Space (Cambridge Univ. Press, Cambridge, 1982).

\bibitem {Unruh-RMP}L.C.B. Crispino, A. Higuchi, G.E.A. Matsas,
The Unruh effect and its applications, Rev. Mod. Phys. \textbf{80} (2008)
787-838.

\bibitem {Intp-book}G. Auletta, Foundations and Interpretation of
Quantum Mechanics (World Scientific, Singapore, 2001).

\bibitem {Fuchs}C.A. Fuchs, Quantum mechanics as quantum information
(and only a little more), Prepint at <http://arXiv.org/quant-ph/0205039>.

\bibitem {i-causality}M. Paw\l owski, T. Paterek, D. Kaszlikowski, V. Scarani,
A. Winter, M. \.{Z}ukowski, Information causality as a physical
principle, Nature \textbf{461} (2009) 1101-1104.

\bibitem {Colbeck-Renner}R. Colbeck, R. Renner, No extension of
quantum theory can have improved predictive power, Nature Commun. \textbf{2} (2011)
411.

\bibitem {Bell}J.S. Bell, On the Einstein-Podolsky-Rosen paradox,
Physics (Long Island City, N.Y.) \textbf{1} (1964) 195-200.

\bibitem {Tipler}F.J. Tipler, Quantum nonlocality does not exist,
Proc. Natl. Acad. Sci. (USA) \textbf{111} (2014) 11281-11286.

\bibitem {ICQT}Z.-B. Chen, The information-complete quantum theory,
Quantum Engineering {\bf 2022} (2022) 9203196.

\bibitem {gGUT}Z.-B. Chen, Synopsis of a unified theory for all forces
and matter, arXiv:1611.02662.

\bibitem {Ashtekar}A. Ashtekar, New variables for classical and
quantum gravity, Phys. Rev. Lett. \textbf{57} (1986) 2244-2247.

\bibitem {Barbero}J.F. Barbero, Real Ashtekar variables for Lorentzian
signature space times, Phys. Rev. D \textbf{51} (1995) 5507-5510.

\bibitem {Immirzi}G. Immirzi, Real and complex connections for
canonical gravity, Class. Quantum Grav. \textbf{14} (1997) L177-L181.

\bibitem {pureEE1}C.H. Bennett, H. Bernstein, S. Popescu, B. Schumacher,
Concentrating partial entanglement by local operations, Phys. Rev. A
\textbf{53} (1996) 2046-2062.

\bibitem {pureEE2}S. Popescu, D. Rohrlich, Thermodynamics and the
measure of entanglement, Phys. Rev. A \textbf{56} (1997) R3319-R3321.

\bibitem {Jacobson}T. Jacobson, Thermodynamics of spacetime: the
Einstein equation of state, Phys. Rev. Lett. \textbf{75} (1995) 1260-1263.

\bibitem {Verlinde}E. Verlinde, On the origin of gravity and the laws
of Newton, JHEP \textbf{04} (2011) 029.

\bibitem {Kiefer}C. Kiefer, Conceptual problems in quantum gravity and
quantum cosmology, ISRN Math. Phys. \textbf{2013} (2013) 509316; arXiv:1401.3578.

\bibitem {eH-Li}H. Li, F.D.M. Haldane, Entanglement spectrum as a
generalization of entanglement entropy: Identification of topological order in
non-Abelian fractional quantum Hall effect states, Phys. Rev. Lett.
\textbf{101} (2008) 010504.

\bibitem {eH-1Dfermi}A.M. Turner, F. Pollmann, E. Berg,
Topological phases of one-dimensional fermions: An entanglement point
of view, Phys. Rev. B \textbf{83} (2011) 075102.

\bibitem {EuA-Hawking}S.W. Hawking, G.T. Horowitz, The
gravitational Hamiltonian, action, entropy and surface terms, Class. Quantum
Grav. \textbf{13} (1996) 1487-1498.

\bibitem {EuA-Pad}S. Kolekar, T. Padmanabhan, Holography in
action, Phys. Rev. D \textbf{82} (2010) 024036.

\bibitem {Preskill}J. Preskill, Quantum Information and Computation
(Lecture Notes for ph219/cs219, CIT, California, 2001).

\bibitem {QND-rmp}V.B. Braginsky, F.Ya. Khalili, Quantum
nondemolition measurements: the route from toys to tools, Rev. Mod. Phys.
\textbf{68} (1996) 1-11.

\bibitem {Page-Wootters}D.N. Page, W.K. Wootters, Evolution without
evolution: Dynamics described by stationary observables, Phys. Rev. D
\textbf{27} (1983) 2885-2892.

\bibitem {QTime}V. Giovannetti, S. Lloyd, L. Maccone, Quantum
time, Phys. Rev. D \textbf{92} (2015) 045033.

\bibitem {Rovelli-eigen}H.A. Morales-T\'{e}cotl, C. Rovelli, Loop
space representation of quantum fermions and gravity, Nucl. Phys. B
\textbf{451} (1995) 325-361.

\bibitem {Spin-Fermi}M. Han, C. Rovelli, Spin-foam fermions: PCT
symmetry, Dirac determinant and correlation functions, Class. Quantum Grav.
\textbf{30} (2013) 075007.

\bibitem {cano-Ferimi}M. Bojowald, R. Das, Canonical gravity with
fermions, Phys. Rev. D \textbf{78} (2008) 064009.

\bibitem {Partovi}M.H. Partovi, Entanglement versus Stosszahlansatz:
Disappearance of the thermodynamic arrow in a high-correlation environment,
Phys. Rev. E \textbf{77} (2008) 021110.

\bibitem {FGP}E. Frodden, A. Ghosh, A. Perez, A local first law
for black hole thermodynamics, Phys. Rev. D \textbf{87} (2013) 121503(R).

\bibitem {Bianchi12}E. Bianchi, Entropy of non-extremal black holes
from loop gravity, arXiv:1204.5122.

\bibitem {Bianchi13}E. Bianchi, Black hole entropy from graviton
entanglement, arXiv:1211.0522.

\bibitem {Rovelli-Unruh}G. Chirco, H.M. Haggard, A. Riello, C. Rovelli,
Spacetime thermodynamics without hidden degrees of freedom, Phys.
Rev. D \textbf{90} (2014) 044044.

\bibitem {Perez}A. Perez, Statistical and entanglement entropy for
black holes in quantum geometry, Phys. Rev. D \textbf{90} (2014) 084015.

\bibitem {Smolin-Unruh}L. Smolin, General relativity as the equation
of state of spin foam, Class. Quantum Grav. \textbf{31} (2014) 195007.

\bibitem {Bekenstein}J.D. Bekenstein, Black holes and the second law,
Nuovo Cimento Lett. \textbf{4} (1972) 737-740.

\bibitem {BekensteinPRD}J.D. Bekenstein, Black holes and entropy,
Phys. Rev. D \textbf{7} (1973) 2333-2346.

\bibitem {Hawking}S.W. Hawking, Black hole explosions, Nature
\textbf{248} (1974) 30.

\bibitem {tHooft}G. 't Hooft, Dimensional reduction in quantum
gravity, Prepint at <http://arXiv.org/gr-qc/9310026>.

\bibitem {Susskind}L. Susskind, The world as a hologram, J. Math.
Phys. \textbf{36} (1995) 6377-6396.

\bibitem {holoRMP}R. Bousso, The holographic principle, Rev. Mod.
Phys. \textbf{74} (2002) 825-874.

\bibitem {cograin}E.R. Livine, From coarse-graining to holography in
loop quantum gravity, arXiv:1704.04067.

\bibitem {cc-Weinberg}S. Weinberg, The cosmological constant problem,
Rev. Mod. Phys. \textbf{61} (1989) 1-23.

\bibitem {mono}B.M. Terhal, Is entanglement monogamous? IBM J. Res.
Dev. \textbf{48} (2004) 71-78.

\bibitem {Bojowald}M. Bojowald, Information loss, made worse by
quantum gravity, arXiv:1409.3157.

\bibitem {MTZ}C.W. Misner, K.S. Thorne, W.H. Zurek, John Wheeler,
relativity, and quantum information, Phys. Today \textbf{67} (April, 2014) 40-46.


\end{thebibliography}
\end{document}